\documentclass{aa}

\usepackage{graphicx}
\usepackage{txfonts}
\usepackage{multirow}

\newcommand{\arcs}{$^{\prime\prime}$}

\begin{document}

\title{Doubly eclipsing systems}

\author{P. Zasche\inst{1},
        D. Vokrouhlick\'y\inst{1},
        M. Wolf\inst{1},
        H. Ku\v{c}\'akov\'a\inst{1},
        J. K\'ara\inst{1},
        R. Uhla\v{r}\inst{2},
        M. Ma\v{s}ek\inst{3},
        Z. Henzl\inst{4}, and
        P. Caga\v{s}\inst{5}}

\titlerunning{Doubly eclipsing systems}
\authorrunning{Zasche et~al.}

\institute{$^{1}$ Astronomical Institute, Charles University, Faculty
                 of Mathematics and Physics, CZ-180~00, Praha 8,
                 V~Hole\v{s}ovi\v{c}k\'ach 2, \\ Czech
                 Republic , \email{zasche@sirrah.troja.mff.cuni.cz}, \\
           $^{2}$ Private Observatory, Poho\v{r}\'{\i} 71, CZ-254~01,
                 J\'{\i}lov\'e u Prahy, Czech Republic, \\
           $^{3}$ Institute of Physics, Czech Academy of Sciences, Na
                 Slovance 1999/2, CZ-182 21 Praha 8, Czech Republic, \\
           $^{4}$ Variable Star and Exoplanet Section, Czech Astronomical
                 Society, \\
           $^{5}$ BSObservatory, Modr\'a 587, CZ-760~01, Zl\'{\i}n, Czech Republic}

\date{Received 17 July 2019; accepted 22 August 2019}

\abstract
{Only several doubly eclipsing quadruple stellar systems are known to date, and
 no dedicated effort to characterize population properties of these interesting objects
 has yet been made.}
{Our first goal was to increase number of known doubly eclipsing systems such that the resulting
 dataset would allow us to study this category of objects via statistical means. In order to minimize
 biases, we used long-lasting, homogeneous, and well-documented photometric surveys. Second, a common
 problem of basically all known doubly eclipsing systems is the lack of  proof that they
 constitute gravitationally bound quadruple system in the 2+2 architecture (as opposed to two
 unrelated binaries that are projected onto the  same location in the sky by chance). When possible,
 we thus sought evidence for the relative motion of the two binaries. In that case, we tried to
 determine the relevant orbital periods and other parameters.}
{We analysed photometric data for eclipsing binaries provided by the OGLE survey and we focused on
 the LMC fields. We found a large number of new doubly eclipsing systems (our discoveries are three
 times more numerous than the previously known cases in this dataset). In order to prove relative
 motion of the binaries about a common  centre of mass, we made use of the fact that OGLE photometry
 covers several years. With a typical orbital period of days for the observed binaries, we sought
 eclipse time variations (ETVs) on the timescale comparable to a decade (this is the same method
 used for an archetype of the doubly eclipsing system, namely V994~Her). In the cases where we
 were able to detect the ETV period, the difference between the inner and outer periods in the
 quadruple system is large enough. This allows us to interpret ETVs primarily as the light-time effect,
  thus providing an interesting constraint on masses of the binaries.}
{In addition to significantly enlarging the database of known doubly eclipsing systems, we performed a
 thorough analysis of $72$ cases. ETVs for $28$ of them (39\% of the studied cases) showed
 evidence of relative motion. Among these individual systems, we note OGLE~BLG-ECL-145467, by far
 the most interesting case; it is bright ($12.6$~mag in I filter), consists of two
 detached binaries with periods of $\simeq 3.3$~d and $\simeq 4.9$~d (making it a candidate for a $3:2$
 resonant system) revolving about each other in only $\simeq 1538$~d. Distribution of the orbital
 period ratio $P_{\rm A}/P_{\rm B}$ of binaries in 2+2 quadruples shows statistically significant
 excess at $\simeq 1$ and $\simeq 1.5$. The former is likely a natural statistical preference in
 weakly interacting systems with periods within the same range. The latter is thought to be
 evidence of a capture in the $3:2$ mean motion resonance of the two binaries. This sets
 important constraints on evolutionary channels in these systems.}
{The total number of doubly eclipsing systems increased to $146$, more than $90$\% of which are at
 low declinations on the southern sky. This motivates us to  use southern hemisphere facilities to
 further characterize these systems, and to seek possibilities to complement this dataset with
 northern sky systems.}


\maketitle


\section{Introduction}

Some stars in the solar  neighbourhood, in our Galaxy, or in neighbouring galaxies  live in
solitude; the majority, however, have companions. They can form binaries, triples, or even higher
multiples, bound gravitationally together and exhibiting motion about their common centre of mass.
Because of their increasing complexity, higher  multiples are  less frequent
\citep[e.g.][]{2018ApJS..235....6T}. Still, analysis of their architecture can bring interesting
results. For instance, the first multiples that arise in various non-trivial architectures are
stellar quadruple systems. Long-term stability dictates that they may exist as a triple system
accompanied by an additional and distant component ($3+1$ variant). However, there is also another
possibility. Quadruples may also exist as a pair of binaries, whose centres of mass revolve about
each other on a nearly elliptic orbit. This is known as  the $2+2$ variant of quadruples. The
census of known quadruple stellar systems is still quite low, but they can bring interesting
insights to our knowledge of  stellar formation, evolution, and the processes acting during the
life of the star \citep[see  e.g.][]{2018AJ....155..160T,2018MNRAS.479.4749B}. As pointed out by
\citeauthor{2018AJ....155..160T}, the origin of the quadruple systems is still an intriguing
question, with even a possibility that their two architecture variants ($3+1$ and $2+2$) form via
different mechanisms and channels. For that reason, further analysis of quadruples is desirable.

Here we focus on systems in $2+2$ architecture, which potentially offer an interesting possibility
when both binaries are eclipsing. This is a very fortuitous situation because the analysis of
classical eclipsing binaries still  generally provides the most precise method of deriving stellar
parameters, such as masses or radii \citep[e.g.][]{2012ocpd.conf...51S}. Several dozens of
candidates for such systems were discovered during the last decade mainly due to the long-lasting
photometric surveys such as the Optical Gravitational Lensing Experiment (OGLE)
(\citealt{2008AcA....58...69U}, and \citealt{2015AcA....65....1U}). They were dubbed
\emph{\emph{`doubly eclipsing systems'}}, or sometimes   `double-eclipsing' systems.%
\footnote{The doubly eclipsing systems, in our concept, should not be confused with
 the rare situation of triples exhibiting exceptional coplanarity, where both the
 inner binary system is eclipsing and the third star also shows mutual eclipses
 with components in the binary \citep[e.g.][and references therein]{borko2019}.
 }
Characteristic to them are two sets of eclipses belonging to the time series of photometry of a
single source on the sky. The very first of these systems, V994~Her, was discovered more then ten
years ago by \cite{2008MNRAS.389.1630L}, and it is also the first that was proven to constitute a
real, gravitationally bound quadruple system, due to determination of mutual motion of the two
binaries about a common centre of mass \citep[see][]{2016AaA...588A.121Z}. V994~Her was soon
followed with OGLE LMC-ECL-10429 by \citet{ofir}, which was the first suggested to have the orbital
periods of the binaries in mutual $3:2$ resonance.
\begin{table*}
 \caption{Optically resolved pairs of eclipsing binaries. Because of their
  recognized separation, we do not include them in the category of doubly eclipsing
  systems.} \label{NonDoublyEclipsing}
\centering 
\begin{tabular}{lccccll}
 \hline\hline\noalign{\smallskip}
    System name                                &          RA            &             DE                               & $P_{\rm A}$ [d] & Type & Distance                          & Reference                   \\
  \hline\noalign{\smallskip}
 \object{BV Dra}                               & 15$^h$11$^m$50$^s$.36  & $+61^\circ$51$^\prime$25$^{\prime\prime}$.3  & 0.35007   & EW   & \multirow{2}{*}{ \bigg\} 16\arcs} & \multirow{2}{*}{\cite{1982AaAS...48...85G}} \\
 \object{BW Dra}                               & 15$^h$11$^m$50$^s$.10  & $+61^\circ$51$^\prime$41$^{\prime\prime}$.2  & 0.29216   & EW   &                                   &                             \\ \hline
 \object{CY~Tri = CzeV337}                     & 02$^h$06$^m$40$^s$.15  & $+33^\circ$43$^\prime$29$^{\prime\prime}$.1  & 0.33343   & EW   & \multirow{2}{*}{ \bigg\} 9\arcs}  & \multirow{2}{*}{New - this study}\\
 \object{CY~Tri = CzeV621}                     & 02$^h$06$^m$40$^s$.69  & $+33^\circ$43$^\prime$23$^{\prime\prime}$.8  & 0.5373    & EA   &                                   &                             \\ \hline
 \object{CzeV513 = 2MASS 05472403+3045225}     & 05$^h$47$^m$24$^s$.01  & $+30^\circ$45$^\prime$22$^{\prime\prime}$.9  & 0.75007   & EA   & \multirow{2}{*}{ \bigg\} 12\arcs} & \multirow{2}{*}{New - this study}\\
 \object{CzeV609 = 2MASS 05472319+3045181}     & 05$^h$47$^m$23$^s$.19  & $+30^\circ$45$^\prime$18$^{\prime\prime}$.2  & $>$10     & EA   &                                   &                             \\ \hline
\multirow{2}{*}{\object{CRTS J065302.9+381408}}& 06$^h$53$^m$03$^s$.03  & $+38^\circ$14$^\prime$08$^{\prime\prime}$.8  & 1.86640   & EA   & \multirow{2}{*}{ \bigg\} 3.5\arcs}& \multirow{2}{*}{\cite{2014ApJS..213....9D}} \\
                                               & 06$^h$53$^m$02$^s$.81  & $+38^\circ$14$^\prime$06$^{\prime\prime}$.2  & 1.24652   & EA   &                                   &                             \\ \hline
\multirow{2}{*}{\object{CoRoT 211659387}}      & 19$^h$04$^m$00$^s$.93  & $+03^\circ$30$^\prime$32$^{\prime\prime}$.2  & 0.39396   & EW   & \multirow{2}{*}{ \bigg\}  13\arcs}& \multirow{2}{*}{\cite{2017MNRAS.471.1230H}} \\
                                               & 19$^h$04$^m$00$^s$.80  & $+03^\circ$30$^\prime$19$^{\prime\prime}$.3  & 4.0005    & EA   &                                   &                             \\ \hline
  \noalign{\smallskip}\hline
\end{tabular}
\end{table*}

For the sake of completeness, we also note optically resolved, separate pairs of eclipsing binaries
very close to each other on the sky (see Table~\ref{NonDoublyEclipsing}). Because of the large
separation, their mutual motion is not determined, but it is conceivable that they form
gravitationally bound quadruples. An archetype for this category is the system consisting of BV Dra
and BW Dra. The system CoRoT 211659387, often classified as doubly eclipsing, was actually resolved
thanks to the PDR \citep{2019CoSka..49..132Z} and the ZTF survey
\citep{2019PASP..131a8003M,2019PASP..131a8002B}. The CRTS survey \citep{2009ApJ...696..870D} helped
to resolve the other systems listed in  Table~\ref{NonDoublyEclipsing}. Using a small telescope, these
systems can be seen to  behave like doubly eclipsing systems, but because of the recognized separation we do not
include them in our  analysis.

Once some level of understanding of the doubly eclipsing system is reached, especially if the
parameters of the mutual motion of the binaries are determined, the quadruple may become a valuable
laboratory for testing important astrophysical tasks. First, in a  traditional fashion, data
for individual binaries in the system can be used to track their evolutionary status, derive their
age, determine their chemical composition, and so on. The added value of these data is
that the two binaries should share most of these same parameters  (age, chemical composition,
distance, etc.). Having thus two independent ways to derive the same parameters should in principle
help to increase their accuracy. Second, the dynamics of $2+2$ quadruples is sufficiently rich and, if
compact enough, even the secular evolution may be on interestingly short timescales \citep[see
e.g.][]{v16,hpz16,hl17}. In addition to  the plethora of short-term effects exhibited by eclipse time
variations (ETVs) \citep[e.g.][]{borko2003,borko2011},   secular effects such as apsidal
precession of inner binaries or inclination change can also be detected
\citep[e.g.][]{2016AaA...588A.121Z,2013ApJ...768...33R,jury18}. Finally, as for any class of
stellar systems, if   data are gathered for a large enough ensemble of cases, we can start
studying statistical trends related to different parameters. Those typically hide valuable hints
about formation mechanisms and/or evolutionary tracks of the binary orbits driven by mutual
gravitational interaction and tides \cite[e.g.][]{2008MNRAS.389..925T}. The $2+2$ quadruples seem
to have a special status in this endeavour as also argued in a recent review by
\citet{2018AJ....155..160T}. All these arguments motivated their study, especially in the situation
when not many doubly eclipsing systems are still not known today.

\section{Data source and analysis strategy} \label{selection}

Observations provided by the OGLE survey \citep{1992AcA....42..253U} are very suitable for our task
of finding new quadruple systems thanks to their completeness, availability, well-documented and easily downloadable catalogue, long-term stable and uniform photometry of the observed
targets, and data covering more than a decade for most of the fields. All these qualities
facilitate our task. We focused on the LMC fields for several reasons: (i) OGLE~III data were
available for a total of eight seasons, OGLE~IV for four seasons, sometimes also complemented by data
from OGLE~II from four seasons, (ii)  MACHO data are also available for some of the
systems, and (iii) southern declinations are suitable for the Danish 1.54m telescope at La Silla,
available to our group, which allows dedicated extensions of the survey data.  We note that OGLE LMC
fields contain  26121 documented eclipsing binaries.%
\footnote{There are even more detected eclipsing binaries (40204) in the OGLE~IV
 data \citep{2016AcA....66..421P}, but we use them only in conjunction with the OGLE~III
 observations  because our primary strategy relied on the long time span of the data.}
This is both large enough to hope for a meaningful search for doubly eclipsing systems
and small enough to be treatable in a semi-manual way by a single person.

\begin{table*}
\caption{Currently known doubly eclipsing systems: sky position given by right ascension (RA)
 and declination (DE) (ordered by increasing value of RA), and fundamental periods
 $P_{\rm A}$ and $P_{\rm B}$ of the two binaries. Also given is  the number of the Group
 for the confirmed quadruples according to our division in Section \ref{results}. } \label{InfoSystems}
 \tiny
  \centering \scalebox{0.7}{
\begin{tabular}{lcccccl|lcccccl}
   \hline\hline\noalign{\smallskip}
    System name               &          RA           &             DE                   & $P_{\rm A}$ [d] & $P_{\rm B}$ [d] & Group& Ref. &  System name                        &         RA           &             DE                     & $P_{\rm A}$ [d]  & $P_{\rm B}$ [d] &Group& Ref.  \\
  \hline\noalign{\smallskip}
  \object{OGLE SMC-ECL-0629}  & 00$^h$43$^m$10$^s$.90 & $-73^\circ$23$^\prime$41$^{\prime\prime}$.0 & 3.95327    & 244.79804 &      & (1)  & \object{OGLE LMC-ECL-21094}         & 05$^h$37$^m$54$^s$.61 & $-69^\circ$40$^\prime$47$^{\prime\prime}$.3 & 3.0101772    & 1.0141844 &  2  & New   \\
  \object{OGLE SMC-ECL-1076}  & 00$^h$46$^m$35$^s$.57 & $-73^\circ$14$^\prime$26$^{\prime\prime}$.8 & 6.40349    & 4.30215   &      & (1)  & \object{OGLE LMC-ECL-21456}         & 05$^h$38$^m$44$^s$.16 & $-69^\circ$05$^\prime$42$^{\prime\prime}$.2 & 1.0817076    & 2.6574721 &     & (4)   \\
  \object{EPIC 220204960}     & 00$^h$48$^m$32$^s$.65 & $+00^\circ$10$^\prime$18$^{\prime\prime}$.5 & 13.2735    & 14.4158   &      & (2)  & \object{OGLE LMC-ECL-21569}         & 05$^h$38$^m$58$^s$.39 & $-69^\circ$04$^\prime$35$^{\prime\prime}$.1 & 1.9815435    & 2.9328514 &  1  & (4)   \\
  \object{OGLE SMC-ECL-1758}  & 00$^h$49$^m$55$^s$.23 & $-73^\circ$16$^\prime$51$^{\prime\prime}$.3 & 0.92917    & 3.73518   &      & (1)  & \object{OGLE LMC-ECL-21603}         & 05$^h$39$^m$04$^s$.89 & $-69^\circ$29$^\prime$37$^{\prime\prime}$.7 & 1.7811269    & 7.118170  &     & New   \\
  \object{OGLE SMC-ECL-2036}  & 00$^h$51$^m$04$^s$.28 & $-72^\circ$47$^\prime$38$^{\prime\prime}$.9 & 1.25371    & 21.75096  &      & (1)  & \object{OGLE LMC-ECL-21991}         & 05$^h$40$^m$05$^s$.90 & $-69^\circ$45$^\prime$05$^{\prime\prime}$.0 & 10.328120    & 18.2655   &     & New   \\
  \object{OGLE SMC-ECL-2141}  & 00$^h$51$^m$24$^s$.57 & $-72^\circ$40$^\prime$15$^{\prime\prime}$.8 & 0.56554    & 1.27330   &      & (1)  & \object{OGLE LMC-ECL-21994}         & 05$^h$40$^m$06$^s$.34 & $-70^\circ$06$^\prime$43$^{\prime\prime}$.9 & 7.4112146    & 6.0404463 &  3  & New   \\
  \object{OGLE SMC-ECL-2208}  & 00$^h$51$^m$39$^s$.69 & $-73^\circ$18$^\prime$45$^{\prime\prime}$.3 & 5.72602    & 2.61777   &      & (1)  & \object{OGLE LMC-ECL-22148}         & 05$^h$40$^m$29$^s$.73 & $-70^\circ$04$^\prime$59$^{\prime\prime}$.9 & 1.8267531    & 2.7147783 &     & New   \\
  \object{OGLE SMC-ECL-2529}  & 00$^h$52$^m$43$^s$.03 & $-72^\circ$42$^\prime$24$^{\prime\prime}$.5 & 1.07455    & 6.54472   &      & (1)  & \object{OGLE LMC-ECL-22159}         & 05$^h$40$^m$31$^s$.25 & $-69^\circ$11$^\prime$33$^{\prime\prime}$.0 & 2.9884068    & 3.4084378 &     & (4)   \\
  \object{OGLE SMC-ECL-2586}  & 00$^h$52$^m$53$^s$.03 & $-73^\circ$11$^\prime$12$^{\prime\prime}$.2 & 1.25169    & 1.51224   &      & (1)  & \object{OGLE LMC-ECL-22281}         & 05$^h$40$^m$49$^s$.55 & $-69^\circ$13$^\prime$05$^{\prime\prime}$.8 & 2.8675292    & 1.503588  &     & New   \\
  \object{OGLE SMC-ECL-2715}  & 00$^h$53$^m$23$^s$.28 & $-72^\circ$37$^\prime$29$^{\prime\prime}$.2 & 0.76321    & 1.02086   &      & (1)  & \object{OGLE LMC-ECL-22434}         & 05$^h$41$^m$11$^s$.85 & $-70^\circ$52$^\prime$00$^{\prime\prime}$.3 & 1.5813927    & 1.4755317 &     & New   \\
  \object{OGLE SMC-ECL-2896}  & 00$^h$54$^m$02$^s$.92 & $-72^\circ$32$^\prime$38$^{\prime\prime}$.0 & 0.65978    & 1.18166   &      & (1)  & \object{OGLE LMC-ECL-22891}         & 05$^h$42$^m$15$^s$.83 & $-69^\circ$04$^\prime$55$^{\prime\prime}$.3 & 0.8755663    & 0.8546343 &     & New   \\
  \object{OGLE SMC-ECL-3284}  & 00$^h$55$^m$42$^s$.92 & $-73^\circ$35$^\prime$37$^{\prime\prime}$.0 & 1.01122    & 2.43480   &      & (1)  & \object{OGLE LMC-ECL-23000}         & 05$^h$42$^m$37$^s$.06 & $-69^\circ$04$^\prime$12$^{\prime\prime}$.4 & 1.8998605    & 1.2455112 &  3  & New   \\
  \object{OGLE SMC-ECL-4418}  & 01$^h$01$^m$51$^s$.55 & $-72^\circ$05$^\prime$49$^{\prime\prime}$.1 & 0.71821    & 3.26509   &      & (1)  & \object{OGLE LMC-ECL-23469}         & 05$^h$44$^m$01$^s$.24 & $-69^\circ$16$^\prime$56$^{\prime\prime}$.1 & 1.3995586    & 1.4808157 &     & New   \\
  \object{OGLE SMC-ECL-4731}  & 01$^h$03$^m$32$^s$.26 & $-72^\circ$02$^\prime$23$^{\prime\prime}$.2 & 0.73811    & 0.61356   &      & (1)  & \object{OGLE LMC-ECL-23823}         & 05$^h$45$^m$27$^s$.71 & $-69^\circ$46$^\prime$22$^{\prime\prime}$.3 & 1.1844660    & 204.424   &     & New   \\
  \object{OGLE SMC-ECL-4908}  & 01$^h$04$^m$35$^s$.34 & $-72^\circ$07$^\prime$41$^{\prime\prime}$.1 & 2.55792    & 2.85180   &      & (1)  & \object{CzeV343}                    & 05$^h$48$^m$24$^s$.01 & $+30^\circ$57$^\prime$03$^{\prime\prime}$.6 & 1.209364     & 0.806869  &     & (5)   \\
  \object{OGLE SMC-ECL-5015}  & 01$^h$05$^m$20$^s$.44 & $-72^\circ$03$^\prime$43$^{\prime\prime}$.4 & 0.76283    & 1.15616   &      & (1)  & \object{OGLE LMC-ECL-25635}         & 06$^h$01$^m$17$^s$.27 & $-69^\circ$03$^\prime$05$^{\prime\prime}$.1 & 5.2254980    & 610.933   &     & New   \\
  \object{V482 Per}           & 04$^h$15$^m$41$^s$.33 & $+47^\circ$25$^\prime$19$^{\prime\prime}$.9 & 2.4467526  & 6.001749  &      & (3)  & \object{CzeV1640}                   & 06$^h$07$^m$18$^s$.39 & $+28^\circ$07$^\prime$25$^{\prime\prime}$.1 & 0.554234     & 0.842581  &     & (6)   \\
  \object{OGLE LMC-ECL-00728} & 04$^h$46$^m$13$^s$.20 & $-69^\circ$03$^\prime$51$^{\prime\prime}$.2 & 1.7705911  & 1.0937835 &      & New  & \object{CoRoT 223993566}            & 06$^h$41$^m$49$^s$.17 & $+10^\circ$07$^\prime$19$^{\prime\prime}$.4 & 1.18067      & 0.934856  &     & (7)   \\
  \object{OGLE LMC-ECL-01050} & 04$^h$48$^m$39$^s$.69 & $-68^\circ$27$^\prime$30$^{\prime\prime}$.9 & 1.6707882  & 26.4766   &      & New  & \object{CoRoT 110829335}            & 06$^h$49$^m$04$^s$.86 & $-05^\circ$51$^\prime$31$^{\prime\prime}$.3 & 8.9304       & 50.3075   &     & (7)   \\
  \object{OGLE LMC-ECL-02156} & 04$^h$53$^m$07$^s$.48 & $-70^\circ$11$^\prime$43$^{\prime\prime}$.6 & 4.5172360  & 1.2614425 & 3    & (4)  & \object{1SWASP J093010.78+533859.5} & 09$^h$30$^m$10$^s$.75 & $+53^\circ$38$^\prime$59$^{\prime\prime}$.8 & 1.3055472    & 0.2277142 &  2  & (8)   \\
  \object{OGLE LMC-ECL-02310} & 04$^h$53$^m$37$^s$.99 & $-69^\circ$13$^\prime$42$^{\prime\prime}$.7 & 1.0463484  & 16.06224  & 2    & New  & \object{OGLE GD-ECL-00259}          & 10$^h$37$^m$26$^s$.39 & $-62^\circ$29$^\prime$00$^{\prime\prime}$.3 & 1.1423587    & 0.6449106 &     & (9)   \\
  \object{OGLE LMC-ECL-02903} & 04$^h$55$^m$12$^s$.82 & $-68^\circ$52$^\prime$24$^{\prime\prime}$.7 & 2.0799677  & 6.5669915 & 1    & (4)  & \object{OGLE GD-ECL-03436}          & 10$^h$49$^m$37$^s$.96 & $-61^\circ$59$^\prime$40$^{\prime\prime}$.6 & 1.6798563    & 3.1251724 &     & (9)   \\
  \object{OGLE LMC-ECL-03611} & 04$^h$57$^m$00$^s$.38 & $-69^\circ$30$^\prime$42$^{\prime\prime}$.9 & 2.1195730  & 1.396880  &      & New  & \object{OGLE GD-ECL-04406}          & 10$^h$53$^m$36$^s$.00 & $-61^\circ$32$^\prime$53$^{\prime\prime}$.6 & 1.5728462    & 1.6757585 &     & (9)   \\
  \object{OGLE LMC-ECL-03906} & 04$^h$57$^m$39$^s$.65 & $-69^\circ$07$^\prime$52$^{\prime\prime}$.8 & 9.9285502  & 10.63576  &      & New  & \object{OGLE GD-ECL-05310}          & 10$^h$56$^m$39$^s$.08 & $-60^\circ$23$^\prime$01$^{\prime\prime}$.0 & 3.3360910    & 1.8336427 &     & (9)   \\
  \object{OGLE LMC-ECL-04236} & 04$^h$58$^m$28$^s$.60 & $-70^\circ$14$^\prime$11$^{\prime\prime}$.4 & 2.4074806  & 2.4602955 & 1    & New  & \object{OGLE GD-ECL-05390}          & 10$^h$56$^m$57$^s$.83 & $-60^\circ$47$^\prime$39$^{\prime\prime}$.9 & 1.7506700    & 4.2500570 &     & (9)   \\
  \object{OGLE LMC-ECL-04465} & 04$^h$59$^m$09$^s$.97 & $-68^\circ$38$^\prime$49$^{\prime\prime}$.4 & 1.4105797  & 143.735   &      & New  & \object{OGLE GD-ECL-05656}          & 10$^h$57$^m$50$^s$.86 & $-61^\circ$37$^\prime$49$^{\prime\prime}$.3 & 2.1541808    & 1.2867213 &     & (9)   \\
  \object{OGLE LMC-ECL-04623} & 04$^h$59$^m$38$^s$.46 & $-69^\circ$25$^\prime$22$^{\prime\prime}$.8 & 1.6422711  &10.6468130 & 1    & New  & \object{OGLE GD-ECL-07057}          & 11$^h$06$^m$34$^s$.42 & $-61^\circ$10$^\prime$09$^{\prime\prime}$.4 & 1.1605585    & 1.9234148 &     & (9)   \\
  \object{OGLE LMC-ECL-06179} & 05$^h$03$^m$37$^s$.35 & $-68^\circ$56$^\prime$22$^{\prime\prime}$.7 & 1.5127562  & 1.227690  &      & New  & \object{OGLE GD-ECL-07157}          & 11$^h$07$^m$45$^s$.07 & $-61^\circ$20$^\prime$56$^{\prime\prime}$.1 & 0.8128751    & 2.6694423 &  1  & (9)   \\
  \object{OGLE LMC-ECL-06331} & 05$^h$03$^m$57$^s$.72 & $-70^\circ$16$^\prime$22$^{\prime\prime}$.6 & 1.0336030  & 1.212532  &      & New  & \object{OGLE GD-ECL-07443}          & 11$^h$31$^m$05$^s$.11 & $-60^\circ$41$^\prime$55$^{\prime\prime}$.9 & 1.7501509    & 1.4512089 &     & (9)   \\
  \object{OGLE LMC-ECL-06538} & 05$^h$04$^m$28$^s$.59 & $-68^\circ$37$^\prime$38$^{\prime\prime}$.6 & 0.7434252  & 0.590837  &      & New  & \object{OGLE GD-ECL-10263}          & 13$^h$26$^m$23$^s$.30 & $-65^\circ$05$^\prime$37$^{\prime\prime}$.0 & 0.4208822    & 0.3787910 &     & (9)   \\
  \object{OGLE LMC-ECL-06595} & 05$^h$04$^m$35$^s$.86 & $-69^\circ$35$^\prime$31$^{\prime\prime}$.5 & 1.1313483  & 1.188998  &      & New  & \object{OGLE GD-ECL-11021}          & 13$^h$32$^m$56$^s$.60 & $-64^\circ$09$^\prime$50$^{\prime\prime}$.6 & 1.1601455    & 3.0600241 &     & (9)   \\
  \object{OGLE LMC-ECL-07329} & 05$^h$06$^m$08$^s$.85 & $-67^\circ$53$^\prime$15$^{\prime\prime}$.8 & 1.7172776  & 206.43    &      & New  & \object{EPIC 212651213}             & 13$^h$55$^m$43$^s$.46 & $-09^\circ$25$^\prime$05$^{\prime\prime}$.9 & 5.07655      & 13.1947   &     & (10)  \\
  \object{OGLE LMC-ECL-07485} & 05$^h$06$^m$30$^s$.34 & $-68^\circ$34$^\prime$51$^{\prime\prime}$.3 & 8.0282937  & 1.4757278 &      & New  & \object{OGLE BLG-ECL-018877}        & 17$^h$28$^m$41$^s$.22 & $-29^\circ$27$^\prime$48$^{\prime\prime}$.0 & 0.6008759    & 1.5565025 &  1  & (11)  \\
  \object{OGLE LMC-ECL-08902} & 05$^h$09$^m$50$^s$.90 & $-68^\circ$52$^\prime$32$^{\prime\prime}$.7 & 2.3378256  & 25.663    &      & New  & \object{OGLE BLG-ECL-019637}        & 17$^h$29$^m$01$^s$.21 & $-29^\circ$29$^\prime$48$^{\prime\prime}$.1 & 0.4011300    & 0.368949  &     & (11)  \\
  \object{OGLE LMC-ECL-08914} & 05$^h$09$^m$52$^s$.41 & $-69^\circ$23$^\prime$45$^{\prime\prime}$.2 & 2.4548301  & 3.9438804 &      & New  & \object{OGLE BLG-ECL-030128}        & 17$^h$33$^m$22$^s$.18 & $-33^\circ$47$^\prime$47$^{\prime\prime}$.9 & 2.2742881    & 1.9120361 &  2  & (11)  \\
  \object{OGLE LMC-ECL-08957} & 05$^h$09$^m$58$^s$.40 & $-70^\circ$28$^\prime$18$^{\prime\prime}$.6 & 1.3388693  & 135.515   &      & New  & \object{OGLE BLG-ECL-061232}        & 17$^h$40$^m$24$^s$.47 & $-27^\circ$43$^\prime$02$^{\prime\prime}$.9 & 0.3791298    & 1.4676043 &  1  & (11)  \\
  \object{OGLE LMC-ECL-09257} & 05$^h$10$^m$40$^s$.37 & $-67^\circ$09$^\prime$40$^{\prime\prime}$.4 & 0.9324693  & 3.59443   &      & New  & \object{OGLE BLG-ECL-088871}        & 17$^h$44$^m$59$^s$.67 & $-23^\circ$42$^\prime$45$^{\prime\prime}$.3 & 3.8779159    & 5.6508216 &  1  & (11)  \\
  \object{OGLE LMC-ECL-09464} & 05$^h$11$^m$11$^s$.47 & $-67^\circ$09$^\prime$51$^{\prime\prime}$.1 & 1.2895788  & 3.697210  &      & New  & \object{OGLE BLG-ECL-089724}        & 17$^h$45$^m$06$^s$.94 & $-23^\circ$49$^\prime$51$^{\prime\prime}$.8 & 3.4925576    & 0.343487  &     & (11)  \\
  \object{OGLE LMC-ECL-10429} & 05$^h$13$^m$43$^s$.02 & $-69^\circ$18$^\prime$37$^{\prime\prime}$.0 & 3.5779357  & 5.3666155 &      & (4)  & \object{OGLE BLG-ECL-093829}        & 17$^h$45$^m$40$^s$.04 & $-22^\circ$38$^\prime$49$^{\prime\prime}$.8 & 3.7452992    & 0.5210858 &     & (11)  \\
  \object{OGLE LMC-ECL-11224} & 05$^h$15$^m$39$^s$.19 & $-70^\circ$05$^\prime$47$^{\prime\prime}$.4 & 1.9450169  & 3.2025101 &      & New  & \object{OGLE BLG-ECL-100363}        & 17$^h$46$^m$33$^s$.05 & $-20^\circ$53$^\prime$30$^{\prime\prime}$.2 & 4.3521616    & 0.5749267 &     & (11)  \\
  \object{OGLE LMC-ECL-12807} & 05$^h$19$^m$33$^s$.99 & $-68^\circ$16$^\prime$24$^{\prime\prime}$.3 & 1.7252223  & 4.85997   &      & New  & \object{OGLE BLG-ECL-103591}        & 17$^h$46$^m$59$^s$.87 & $-36^\circ$30$^\prime$59$^{\prime\prime}$.2 & 2.2321488    & 2.2833714 &  1  & (11)  \\
  \object{OGLE LMC-ECL-12857} & 05$^h$19$^m$43$^s$.23 & $-68^\circ$38$^\prime$48$^{\prime\prime}$.7 & 2.0826965  & 1.825429  &      & (4)  & \object{OGLE BLG-ECL-104219}        & 17$^h$47$^m$05$^s$.15 & $-35^\circ$02$^\prime$05$^{\prime\prime}$.3 & 0.4683403    & 0.457687  &     & (11)  \\
  \object{OGLE LMC-ECL-13221} & 05$^h$20$^m$38$^s$.71 & $-68^\circ$52$^\prime$26$^{\prime\prime}$.3 & 1.4164210  & 0.7325898 &      & New  & \object{OGLE BLG-ECL-133521}        & 17$^h$50$^m$54$^s$.79 & $-21^\circ$34$^\prime$01$^{\prime\prime}$.2 & 1.0472674    & 1.0388721 &     & (11)  \\
  \object{OGLE LMC-ECL-13737} & 05$^h$21$^m$54$^s$.94 & $-67^\circ$54$^\prime$20$^{\prime\prime}$.6 & 1.2743786  & 6.06765   &      & New  & \object{OGLE BLG-ECL-145467}        & 17$^h$52$^m$05$^s$.58 & $-29^\circ$19$^\prime$43$^{\prime\prime}$.5 & 3.3049105    & 4.9097045 &  1  & (11)  \\
  \object{OGLE LMC-ECL-14370} & 05$^h$23$^m$15$^s$.48 & $-69^\circ$54$^\prime$33$^{\prime\prime}$.7 & 1.0593788  & 1.903334  &      & New  & \object{OGLE BLG-ECL-165082}        & 17$^h$53$^m$45$^s$.95 & $-29^\circ$12$^\prime$59$^{\prime\prime}$.4 & 0.9599463    & 1.092108  &     & (11)  \\
  \object{OGLE LMC-ECL-14375} & 05$^h$23$^m$16$^s$.29 & $-70^\circ$10$^\prime$06$^{\prime\prime}$.9 & 1.0711547  & 73.5596   &      & New  & \object{OGLE BLG-ECL-187370}        & 17$^h$55$^m$38$^s$.24 & $-28^\circ$15$^\prime$49$^{\prime\prime}$.7 &11.9634963    & 87        &     & (11)  \\
  \object{OGLE LMC-ECL-15301} & 05$^h$25$^m$31$^s$.29 & $-69^\circ$26$^\prime$13$^{\prime\prime}$.5 & 0.7373025  & 4.898650  &      & New  & \object{OGLE BLG-ECL-190427}        & 17$^h$55$^m$53$^s$.91 & $-22^\circ$59$^\prime$51$^{\prime\prime}$.1 & 0.9449826    & 2.5137669 &  1  & (11)  \\
  \object{OGLE LMC-ECL-15607} & 05$^h$26$^m$15$^s$.06 & $-69^\circ$04$^\prime$57$^{\prime\prime}$.7 & 1.0479561  & 0.4341706 & 2    & New  & \object{OGLE BLG-ECL-197015}        & 17$^h$56$^m$27$^s$.27 & $-27^\circ$47$^\prime$22$^{\prime\prime}$.8 & 0.3759299    & 6.53287   &     & (11)  \\
  \object{OGLE LMC-ECL-15674} & 05$^h$26$^m$22$^s$.97 & $-68^\circ$49$^\prime$27$^{\prime\prime}$.1 & 1.4332330  & 1.3875757 &      & (4)  & \object{OGLE BLG-ECL-200747}        & 17$^h$56$^m$46$^s$.00 & $-31^\circ$13$^\prime$59$^{\prime\prime}$.1 &42.7652100    & 0.287215  &     & (11)  \\
  \object{OGLE LMC-ECL-15742} & 05$^h$26$^m$33$^s$.10 & $-69^\circ$09$^\prime$13$^{\prime\prime}$.8 & 0.8584714  & 2.1258373 & 2    & (4)  & \object{OGLE BLG-ECL-246147}        & 18$^h$00$^m$43$^s$.37 & $-26^\circ$23$^\prime$20$^{\prime\prime}$.7 & 2.0615552    & 2.1689469 &     & (11)  \\
  \object{OGLE LMC-ECL-16532} & 05$^h$28$^m$06$^s$.72 & $-69^\circ$01$^\prime$11$^{\prime\prime}$.7 & 0.7437700  & 77.91193  & 3    & New  & \object{OGLE BLG-ECL-250817}        & 18$^h$01$^m$07$^s$.45 & $-28^\circ$48$^\prime$31$^{\prime\prime}$.6 & 9.2515062    & 0.4396163 &     & (11)  \\
  \object{OGLE LMC-ECL-16539} & 05$^h$28$^m$08$^s$.74 & $-70^\circ$21$^\prime$11$^{\prime\prime}$.6 & 12.1675847 & 4.539704  &      & (4)  & \object{OGLE BLG-ECL-251128}        & 18$^h$01$^m$09$^s$.24 & $-27^\circ$42$^\prime$44$^{\prime\prime}$.5 & 0.3786368    & 0.406083  &     & (11)  \\
  \object{OGLE LMC-ECL-16549} & 05$^h$28$^m$09$^s$.41 & $-69^\circ$45$^\prime$28$^{\prime\prime}$.6 & 164.78964  & 0.8180397 &      & (4)  & \object{OGLE BLG-ECL-272587}        & 18$^h$03$^m$09$^s$.08 & $-28^\circ$55$^\prime$17$^{\prime\prime}$.0 & 1.1199022    & 3.3785450 &     & (11)  \\
  \object{OGLE LMC-ECL-16831} & 05$^h$28$^m$40$^s$.73 & $-68^\circ$41$^\prime$36$^{\prime\prime}$.8 & 2.1848709  & 918.83    &      & New  & \object{OGLE BLG-ECL-274234}        & 18$^h$03$^m$18$^s$.04 & $-28^\circ$13$^\prime$58$^{\prime\prime}$.4 & 6.5352552    & 91.39     &     & (11)  \\
  \object{OGLE LMC-ECL-16988} & 05$^h$29$^m$00$^s$.39 & $-68^\circ$54$^\prime$34$^{\prime\prime}$.8 & 9.8184680  & 1.4832487 &      & (4)  & \object{OGLE BLG-ECL-277539}        & 18$^h$03$^m$36$^s$.11 & $-28^\circ$07$^\prime$41$^{\prime\prime}$.4 & 0.3753292    & 0.5779823 &     & (11)  \\
  \object{OGLE LMC-ECL-17182} & 05$^h$29$^m$22$^s$.89 & $-68^\circ$44$^\prime$50$^{\prime\prime}$.2 & 2.2372570  & 2.4707908 & 1    & (4)  & \object{OGLE BLG-ECL-282858}        & 18$^h$04$^m$05$^s$.15 & $-32^\circ$20$^\prime$21$^{\prime\prime}$.0 & 0.3992092    & 0.539641  &     & (11)  \\
  \object{OGLE LMC-ECL-17347} & 05$^h$29$^m$44$^s$.76 & $-68^\circ$28$^\prime$49$^{\prime\prime}$.4 & 1.9224015  & 445.85    &      & New  & \object{OGLE BLG-ECL-335648}        & 18$^h$09$^m$24$^s$.52 & $-27^\circ$54$^\prime$21$^{\prime\prime}$.2 & 4.6922359    & 2.7201310 &  2  & (11)  \\
  \object{OGLE LMC-ECL-17637} & 05$^h$30$^m$20$^s$.01 & $-69^\circ$46$^\prime$20$^{\prime\prime}$.8 & 1.5995666  & 0.805298  &      & New  & \object{OGLE BLG-ECL-352722}        & 18$^h$11$^m$23$^s$.20 & $-28^\circ$59$^\prime$00$^{\prime\prime}$.2 & 0.5866713    & 3.28423   &     & (11)  \\
  \object{OGLE LMC-ECL-17913} & 05$^h$30$^m$53$^s$.59 & $-71^\circ$07$^\prime$58$^{\prime\prime}$.4 & 1.3184970  & 2.2143375 & 3    & New  & \object{OGLE BLG-ECL-394187}        & 18$^h$16$^m$50$^s$.47 & $-28^\circ$43$^\prime$30$^{\prime\prime}$.7 & 5.5976216    & 1.1349090 &     & (11)  \\
  \object{OGLE LMC-ECL-17996} & 05$^h$31$^m$04$^s$.05 & $-69^\circ$09$^\prime$29$^{\prime\prime}$.0 & 1.5497505  & 1.7005157 &      & (4)  & \object{OGLE BLG-ECL-398110}        & 18$^h$17$^m$26$^s$.65 & $-26^\circ$30$^\prime$20$^{\prime\prime}$.7 & 1.1164591    & 8.2388401 &  2  & (11)  \\
  \object{OGLE LMC-ECL-18618} & 05$^h$32$^m$20$^s$.12 & $-69^\circ$00$^\prime$06$^{\prime\prime}$.3 & 3.6435437  & 1.6271319 &      & (4)  & \object{OGLE BLG-ECL-403022}        & 18$^h$18$^m$19$^s$.52 & $-24^\circ$44$^\prime$57$^{\prime\prime}$.8 & 2.6422384    & 1.1806143 &     & (11)  \\
  \object{OGLE LMC-ECL-18860} & 05$^h$32$^m$56$^s$.45 & $-67^\circ$55$^\prime$21$^{\prime\prime}$.1 & 2.8344662  & 10.64775  &      & New  & \object{OGLE BLG-ECL-406204}        & 18$^h$18$^m$56$^s$.61 & $-28^\circ$00$^\prime$37$^{\prime\prime}$.1 & 0.5740634    & 1.7556268 &     & (11)  \\
  \object{OGLE LMC-ECL-18966} & 05$^h$33$^m$09$^s$.53 & $-69^\circ$27$^\prime$19$^{\prime\prime}$.8 & 2.8618781  & 1.2371204 &      & New  & \object{V994 Her}                   & 18$^h$27$^m$45$^s$.90 & $+24^\circ$41$^\prime$51$^{\prime\prime}$.0 & 2.0832658    & 1.4200395 &  1  & (12)  \\
  \object{OGLE LMC-ECL-19771} & 05$^h$35$^m$03$^s$.46 & $-68^\circ$49$^\prime$02$^{\prime\prime}$.4 & 2.2344535  & 13.05004  &      & New  & \object{CoRoT 310266512}            & 18$^h$31$^m$19$^s$.74 & $-05^\circ$49$^\prime$54$^{\prime\prime}$.6 & 7.421        & 3.266     &     & (13)  \\
  \object{OGLE LMC-ECL-19852} & 05$^h$35$^m$13$^s$.80 & $-69^\circ$02$^\prime$16$^{\prime\prime}$.2 & 2.0754453  & 1.8847946 &      & (4)  & \object{CoRoT 310284765}            & 18$^h$33$^m$51$^s$.26 & $-05^\circ$39$^\prime$23$^{\prime\prime}$.5 & 2.371125     & 1.8754    &     & (7)   \\
  \object{OGLE LMC-ECL-19896} & 05$^h$35$^m$19$^s$.93 & $-69^\circ$20$^\prime$43$^{\prime\prime}$.5 & 1.4449138  & 1.3599000 &      & New  & \object{EPIC 219217635}             & 18$^h$59$^m$00$^s$.62 & $-17^\circ$15$^\prime$57$^{\prime\prime}$.1 & 3.59486      & 0.61815   &     & (14)  \\
  \object{OGLE LMC-ECL-19942} & 05$^h$35$^m$27$^s$.21 & $-69^\circ$41$^\prime$13$^{\prime\prime}$.7 & 4.8175154  & 1.8836875 &      & New  & \object{KIC 3832716}                & 19$^h$01$^m$34$^s$.6  & $+38^\circ$54$^\prime$17$^{\prime\prime}$.7 & 1.1418769    & 2.1702736 &     & (15)  \\
  \object{OGLE LMC-ECL-20145} & 05$^h$35$^m$52$^s$.22 & $-69^\circ$22$^\prime$44$^{\prime\prime}$.0 & 6.1196213  & 3.7219806 &      & New  & \object{CoRoT 211625668}            & 19$^h$01$^m$50$^s$.75 & $+03^\circ$18$^\prime$28$^{\prime\prime}$.7 & 1.771922     & 5.257641  &     & (16)  \\
  \object{OGLE LMC-ECL-20147} & 05$^h$35$^m$52$^s$.86 & $-68^\circ$54$^\prime$49$^{\prime\prime}$.0 & 2.7578408  & 2.6993979 &      & New  & \object{KIC 4247791}                & 19$^h$08$^m$39$^s$.56 & $+39^\circ$22$^\prime$37$^{\prime\prime}$.0 & 4.100871     & 4.049732  &     & (17)  \\
  \object{OGLE LMC-ECL-20382} & 05$^h$36$^m$20$^s$.30 & $-69^\circ$12$^\prime$32$^{\prime\prime}$.3 & 1.8146366  & 2.46648   &      & New  & \object{HD 181469}                  & 19$^h$18$^m$58$^s$.22 & $+39^\circ$16$^\prime$01$^{\prime\prime}$.8 & 8.653$^A$    & 94.226$^A$&     & (18)  \\
  \object{OGLE LMC-ECL-20901} & 05$^h$37$^m$28$^s$.99 & $-69^\circ$23$^\prime$34$^{\prime\prime}$.4 & 1.5544620  & 6.602784  &      & New  & \object{TYC 3929-724-1}             & 19$^h$24$^m$55$^s$.82 & $+57^\circ$04$^\prime$08$^{\prime\prime}$.4 & 4.10846      & 4.67547   &  2  & New   \\
  \object{OGLE LMC-ECL-20903} & 05$^h$37$^m$29$^s$.21 & $-69^\circ$08$^\prime$41$^{\prime\prime}$.6 & 1.6831082  & 4.2565168 &      & New  & \object{TYC 2693-926-1}             & 20$^h$26$^m$43$^s$.83 & $+35^\circ$20$^\prime$30$^{\prime\prime}$.0 & 1.350447     & 1.099203  &     & (19)  \\
  \object{OGLE LMC-ECL-20932} & 05$^h$37$^m$33$^s$.29 & $-69^\circ$24$^\prime$25$^{\prime\prime}$.0 & 3.8304229  & 1.4679165 & 2    & (4)  & \object{NSVS 154567}                & 22$^h$31$^m$41$^s$.94 & $+68^\circ$46$^\prime$22$^{\prime\prime}$.2 & 11.4838      & 2.93956   &     & (19)  \\
 \noalign{\smallskip}\hline
\end{tabular}} \\
\scriptsize Note: [A] - This is a quintuple system with four different sets of
 eclipses. Therefore, the identification of the two periods $P_{\rm A}$ and $P_{\rm B}$ is
  problematic in this case. In the last column Ref., the following references were
 abbreviated: (1) - \cite{2013AcA....63..323P}, (2) - \cite{2017MNRAS.467.2160R}, (3) -
 \cite{2017ApJ...846..115T}, (4) - \cite{2011AcA....61..103G}, (5) - \cite{2012AaA...544L...3C},
 (6) - \cite{2019RNAAS...3f..80C}, (7) - \cite{2017MNRAS.471.1230H}, (8) -
 \cite{2015AaA...578A.103L}, (9) - \cite{2013AcA....63..115P}, (10) - \cite{2016MNRAS.462.1812R},
 (11) - \cite{2016AcA....66..405S}, (12) - \cite{2016AaA...588A.121Z}, (13) -
 \cite{2015PASP..127..421F}, (14) - \cite{2018MNRAS.478.5135B}, (15) - \cite{2018ApaSS.363..267F},
 (16) - \cite{2012AaA...539A..14E}, (17) - \cite{2012AaA...541A.105L}, (18) -
 \cite{2017AaA...602A..30H}, (19) - \cite{2018PZ.....38....3K}.
\end{table*}

A simple code was written to scan the entire OGLE III LMC dataset, and each source (eclipsing
binary) was considered keeping in mind several criteria: (i) a sufficient number of observations, (ii)
magnitude $< 19$~mag in the I filter, (iii) period of eclipse in the range%
\footnote{We mainly focused on Algol-type detached binaries, hence the shorter periods
 were omitted.}
$0.7 < P_{\rm A} < 15$~d, and (iv) the amplitude of variation larger than its scatter. Using this
strategy, we browsed the OGLE binaries one by one, and we  judged by eye whether the source should be
considered as a potential candidate for a doubly eclipsing system or not. The human eye is a good
instrument for this strategy because a normal eclipsing binary should have all its data points
located tightly  around a well-defined  light curve (LC) profile when phase-folded. Doubly eclipsing systems, on the other hand, should have many deviating data points
outlying from the standard shape of the binary light curve (corresponding to dimming of the second
binary in the system when it occurs in eclipse).

It took us about a month    to search  through the whole catalogue with this procedure,
and this resulted in a list of suspected (candidate) systems. These potential doubly eclipsing
binaries ($156$ in total) were further analysed in detail and many false positives were found. For
some of these systems the second period was detectable even in the complete raw data; for others the second period could only be derived on the residuals after subtraction of the main curve of
pair A. Most often the prominent period belongs to  pair A, and it also has larger amplitude of
photometric variations. This period of pair A ($P_{\rm A}$) was taken from the original OGLE catalogue
by \cite{2016AcA....66..405S}. After completing this scrutiny, we were left with $72$ confirmed
doubly eclipsing systems for which we were able to determine both eclipsing periods. Among them,
$17$ cases had previously been  found to be doubly eclipsing   and were flagged this way in the
original OGLE~III catalogue%
\footnote{The authors of the OGLE catalogue   labelled these systems
 `A blend with other eclipsing binary?' in their \texttt{remarks.txt} documentation file.}
by \cite{2011AcA....61..103G}. This is encouraging and tells us that our procedure is quite
efficient, in fact more successful than the previous automated analysis of the OGLE team \citep{2010AcA....60..109G}.
Obviously, some limitations come from our initial criteria constraining periods and magnitudes of
targets. Our final list of presently known doubly eclipsing systems is given in
Table~\ref{InfoSystems}. We note that it also contains systems from other sources than analysed in this
paper.

\section{Determination of quadruple nature for candidate systems} \label{model}

A double-period source on the sky may not necessarily imply that it represents a quadruple system in the
$2+2$ architecture. In some cases, it may perhaps represent a binary with one or both components
eclipsing. Therefore, our final step towards proving we  deal with the quadruple systems represents
the detection of mutual motions of the two binaries about a common centre of mass. As mentioned in the
Introduction, this has been already achieved for a few systems, but this sample remains very small.

There are several possibilities for achieving the task. For instance, one could rely on a
high-quality  spectroscopic measurements of the candidate systems. If things go right, one could
disentangle spectral lines from the system into two binaries, determine their radial velocities,
and thus detect and characterize their relative motion. Obviously, this assumes spectroscopy
available at least over the whole cycle of the relative orbit. As a rule of thumb, stability of the
$2+2$ systems requires at the zero approximation $P_{\rm AB}\gtrsim 5\,{\rm max}(P_{\rm A},P_{\rm
B})$, with more detailed criteria involving masses of the binaries and eccentricity of the relative
orbit \citep[e.g.][]{ma01}. However, for most systems, $P_{\rm AB}$ is significantly longer and may
extend to years, decades, or even centuries
\citep[e.g.][]{1991A&A...248..485D,2008MNRAS.389..925T,2010ApJS..190....1R}. Moreover, many of the
candidate systems are rather faint, so  middle- or large-scale telescopes would be needed for a
sufficiently good spectroscopic observation. Therefore, we believe spectroscopy is important and
needed for further characterization of already recognized quadruples, but it is not primarily
suitable for their detection.

Because the primary recognition of the system as doubly eclipsing relies entirely on its
photometry, it would have been profitable to use the photometric series to also determine
parameters of their relative orbit. Here again, there are several possibilities. For instance, if
the system is not entirely coplanar (i.e. if the orbital plane of the binaries does not coincide exactly
with the orbital plane of the relative orbit), mutual interaction triggers a precession about a
conserved total angular momentum \citep[typically dominated by the contribution from the relative
orbit; e.g.][]{v16}. This makes the binary orbital planes change their geometry with respect to
the observer and thus affects their light curve. An obvious consequence is a change in depth of the
eclipses. However, the characteristic secular precession period is typically long:  $\propto
P_{\rm AB}^2/{\rm max}(P_{\rm A},P_{\rm B})$ \citep{1975A&A....42..229S}. Unless the system is quite compact, the timescale
needed to reveal some changes in the depth of eclipses is long and, additionally, not very suitable
for determining parameters of the mutual orbit. For that reason, so far we have only one example of a
doubly eclipsing system where changes in the binary inclination have been detected
\citep{2018PASP..130e4204H}.

Narrowing the strategy, we see that the minimum approach (and  at the same time the optimum approach)  will be to
detect a phenomenon having a period equal to the period $P_{\rm AB}$ of the relative orbit. The
most straightforward approach consists of detecting   ETVs, if possible
simultaneously for both binaries A and B. ETVs arise from period changes in both binaries apparent
to the observer, and may have various origins. The most readable is simply due to near-elliptic
motion of the two binaries about the centre of mass. Assuming that the inclination $i_{\rm AB}$ of the
relative orbit is not zero, the systems move periodically towards and away from the observer. The
finite velocity of light then simply affects the periodicity of eclipse series, a phenomenon called
the LIght Time Effect (LITE) \citep[e.g.][]{1959AJ.....64..149I,1990BAICz..41..231M}. An important
aspect of LITE in quadruples is that the ETVs for both binaries are strictly anticorrelated in
time, a property which can be readily checked. The LITE approach has been successfully used to
determine the quadruple nature of the first doubly eclipsing system V994~Her
\citep[e.g.][]{2016AaA...588A.121Z}, a study which may be taken as a proof of concept of our
method. We note that ETVs may also have a different, or additional, origin than simply LITE. We discuss this topic in
more detail  in Sect.~\ref{systems}.

Our model thus assumes motion of all stars in the quadruple composed of three elliptic orbits, both
binaries and the relative orbit. Ephemerides of the eclipsing binaries depend on a  minimum of  two
parameters:%
\footnote{Obviously, the orbits are fully described by six parameters; here we mention only those that can  be derived from our photometric dataset, and the approach we use.}
(i) the epoch JD$_0$ of a reference primary eclipse and (ii) the orbital period $P$. If the orbit
is eccentric, additional parameters are (iii) the eccentricity $e$, (iv) the argument of
pericentre $\omega$, and (v) in some cases also the apsidal motion%
\footnote{Conceptually, the apsidal motion is  revealed as a part
 of ETVs of the respective binary \citep[e.g.][]{2015MNRAS.448..946B}. For reasons of
tradition, we treat this part independently from other ETVs and together with
 determination of the orbital parameters of the binaries.}
$\dot\omega$. This represents altogether ten solved-for parameters of the binary orbits.
Additionally, we consider six parameters of the relative orbit fitted by the ETV time series of
both binaries: (i) the period $P_{\rm AB}$, (ii) the epoch of pericentre $T_0$, (iii) the
amplitudes $A_{\rm A}$ and $A_{\rm B}$ of the ETV series of binary A and B, (iv) the eccentricity
$e_{\rm AB}$, and (v) the argument of pericentre $\omega_{\rm AB}$. Each of these parameter sets is
fitted using standard chi-square methods independently. As a result, we obtain full correlation
analysis, including parameter uncertainties, for the individual sets of the parameters, but we do
not solve correlations across them.

Independently, the light curves of both eclipsing binaries were analysed using the program PHOEBE
\citep{2016ApJS..227...29P}. The output provides individual inclination values $i_{\rm A}$ and
$i_{\rm B}$ for the orbital planes of the two binaries, relative radii of the components, and their
luminosity ratios, among other values. The light curve templates were also used to derive the
epochs of the eclipses using the Automatic Fitting Procedure AFP method introduced earlier
\citep[see][]{2014A&A...572A..71Z}, and those used to  construct the ETV series. This approach also
provides an estimate of uncertainty of the ETV values.

As above, our primary source of the photometric data are the catalogues of different phases of the
OGLE survey. The most interesting systems were also observed by us using the Danish 1.54m telescope
at La Silla during the 2018 and 2019 seasons. Because the OGLE~III data terminated in 2015, our
observations often suitably extended the database to solve for typically long $P_{\rm AB}$ periods.
Additionally, the MACHO photometry \citep{2007AJ....134.1963F} was also used for particular systems
when available, and in some rare cases we also added data from the ASAS-SN database
\citep[e.g.][]{2014ApJ...788...48S,2017PASP..129j4502K}.


\section{Landscape of our main results} \label{results}

We analysed $72$ systems selected and described in Sect.~\ref{selection}, a subset of all
doubly eclipsing sources listed in   Table~\ref{InfoSystems}. In $28$ cases ($39$\% of the
analysed sample) we were able to find evidence of the relative motion of the binaries. According to
the quality of the solution, the results can be divided into three separate groups:
\begin{itemize}
\item  \emph{Group 1} comprises  the systems where our analysis was positively
 conclusive (the  ETV series for both binaries A and B  are well covered with sufficiently
 high signal-to-noise ratio; the  LITE assumption is well justified, which implies a good
 anticorrelation of the ETVs; and the period $P_{\rm AB}$ is well determined, typically
 shorter than the data time span);
\item  \emph{Group 2} comprises  the systems where some indication of period
 variation for both pairs was found, but the results are still not very conclusive
 (the $P_{\rm AB}$ is too long, the signal-to-noise ratio of the ETVs is low, or there is some
 other disturbance of a good solution of parameters of the relative orbit);
\item  \emph{Group 3} comprises  the systems where only  weak or indirect evidence of
 mutual motion of binaries A and B was found (the ETVs were reliable for one
 of the binaries only or   some other effects of a more complicated nature that cannot be explained
 by our simple model were present in the data).
\end{itemize}
The relative fraction of systems in these groups is as follows:   Group 1 contains $46$\% of
the cases,   Group 2 contains $36$\% of the cases, and Group 3 contains $18$\% of the cases. The
overall level of our success rate in proving the doubly eclipsing systems to be true quadruples is
further discussed and interpreted in Sect.~\ref{concl}.

What is the main limitation of our method that causes our effort to sometimes fail? Apart from the
obvious case of a too long period $P_{\rm AB}$, a very important role is played by the amplitudes
$A_{\rm A}$ and $A_{\rm B}$ of the ETV series. We recall that our procedure requires fitting both
series of ETVs for binary A and B: if one of them is not detectable in the photometric noise, for
instance due to   unbalanced masses in the two binaries, our method fails. Additionally, we note
that ETVs are not directly observed quantities, but instead are  constructed. For that reason their
amplitude should be compared with the orbital period of the eclipsing binary itself. As an example,
we assume an ETV  amplitude of $\simeq 0.01$~d (typical to our systems). For a system with an
orbital period of $\simeq 1$~day, such an ETV amplitude represents about $1$\%. This is still
acceptable because we are often able to determine the eclipse epoch at this level of precision.
However, if the binary period were $10$~d or longer, the ETV amplitude would become $\lesssim
0.1$\%, and it would   be problematic to resolve them. So there is also a selection bias in our
approach towards systems with binaries having periods that are not too long  (see
Table~\ref{TabLITE} below).

Once successful, though, the analysis of the two ETV series brings an interesting result. Assuming that
LITE dominates ETVs (see further discussion in Sect.~\ref{systems}) the factor expressing
their anticorrelation  directly provides the ratio of masses $m_{\rm A}$ and $m_{\rm B}$ of the two
binaries. In particular, the ratio of their amplitudes is directly related to the mass ratio
$q_{\rm AB}$:
\begin{equation}
 \frac{A_{\rm B}}{A_{\rm A}} = \frac{m_{\rm A}}{m_{\rm B}} = q_{\rm AB}\; .
\end{equation}
We note that this relation holds independently of $i_{\rm AB}$, provided that its value is not too low to
still assure dominance of LITE in ETVs. The possibility to derive this mass ratio from photometry
only, without spectroscopic observations, is   unique. As mentioned above, many of our systems
are faint (distant or even located in other galaxies), and obtaining good-quality spectroscopy
would require time on medium- to large-scale telescopes.

Remaining ambitious, we note that under favourable circumstances we could do even more. Analysis of the
phase-folded light curves of the two binaries may serve to derive their photometric mass ratios.
This may be   problematic for well-detached systems, but for those with  significant
out-of-eclipse variations and obvious ellipsoidal variations this method can provide us with a good
estimate of the inner mass ratios $q_{\rm A}$ and $q_{\rm B}$. Having now both the outer and inner
mass ratios of the component in the quadruple system, the only missing piece of information for
deriving all individual masses is the total mass of the system.

Unfortunately, this last step is tricky. The fit of ETVs of the binaries itself helps to calculate
the mass function: for instance, for binary A we obtain $m_{\rm B}\sin i_{\rm AB}/ m_{\rm
AB}^{2/3}$, and similarly for  system B. This allows us to determine the minimum masses in the two
systems unless some additional information is known about $i_{\rm AB}$. On the other hand, the
maximum masses can be inferred from the general astrophysical knowledge for stars of a given
luminosity. If the gap between the minimum and maximum values is narrow, the information may be
interesting. In general, however, we need additional information about the nature of the stars
(e.g. from spectroscopy), of the orbit, and thus of $i_{\rm AB}$ (e.g. from interferometry).

\section{Individual systems} \label{systems}

Because of the  heterogeneous sample of systems found as quadruples, we will focus in more
detail only on those with conclusive solutions (stars from Group~1).
The remaining systems will be  briefly mentioned later. The quadruples from  Group~1
have the  final solutions of   their light curves and LITEs given in Tables~\ref{TabLC} and
\ref{TabLITE}. Some of the remaining systems are listed in Table~\ref{OtherSystems}, where we
provide comments on their LITE fits.
\begin{figure}
 \centering 
 \includegraphics[width=0.49\textwidth]{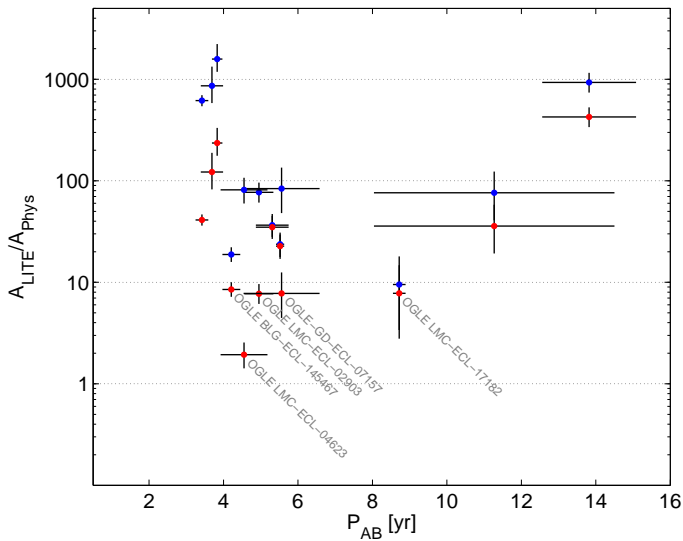}
 \caption{The ratio $A_{\rm LITE}/A_{\rm PD}$ of estimated amplitudes for the
  light-time and physical delay contributions to ETVs for the best-characterized
  systems in Tables~\ref{TabLC} and \ref{TabLITE}: blue symbols for binary A,
  red symbols for binary B (those ones having the ratio below 10 are denoted with
  their names). For sake of simplicity we assumed $i_{\rm AB}\simeq
  90^\circ$, while the other parameters were taken from Table~\ref{TabLITE}. The horizontal
  uncertainty reflects directly uncertainty of $P_{\rm AB}$, the vertical uncertainty
  is a statistical composition of other parameters' uncertainties.}
 \label{LITEvsPhysical}
\end{figure}

In passing we also justify our assumption about LITE dominance in the ETV signal for systems in
Group~1. A competing effect, with the same orbital period $P_{\rm AB}$ of binaries A and B about
their common centre of mass, is named physical delay (PD) by \citet{2013ApJ...768...33R}. PD
originates from a direct mutual perturbation between systems A and B, namely system B changing the
instantaneous mean motion of components in A and vice versa. This change depends on the
instantaneous distance of the two systems, and thus it is modulated during the outer orbital cycle
provided the relative orbit of A and B is eccentric. Because of their different physical natures,
LITE and PD have different dependences on the parameters of the quadruple system, and this allows
us to distinguish between them. In brief, assuming   that the values of eccentricity and
inclination for the outer orbit  are not extreme and that all masses of the stars are comparable,
LITE typically dominates for wide systems with high enough values of  $P_{\rm AB}$, therefore low
values of $P_{\rm A}/P_{\rm AB}$ and $P_{\rm B}/ P_{\rm AB}$. This is true because the LITE
amplitude scales as
\begin{equation}
  A_{\rm LITE}\propto (m_\star/ m_{\rm AB}^{2/3})(\sin i_{\rm AB}/c)(P_{\rm AB}/2\pi)^{2/3} G^{1/3}\; ,
\end{equation}
(where $m_\star$ is the mass of the
companion binary, $m_{\rm AB}$ is the total mass of all stars, $i_{\rm AB}$ is the inclination of
the outer orbit, and $c$ is the light velocity), while the PD amplitude scales as
\begin{equation}
 A_{\rm PD}\propto (3/2\pi) (m_\star/m_{\rm AB})(P^2/P_{\rm AB})\, (1-e^2_{\rm AB})^{-3/2}\, e_{\rm AB} \; ,
\end{equation}
(where $P$ is the orbital period of the system whose ETVs are computed --A or B -- and $e_{\rm
AB}$ is the eccentricity of the outer orbit, the estimate being valid to the linear order in
$e_{\rm AB}$ and for near-coplanar systems, otherwise see Sect.~4.1.2 in
\citealt{2013ApJ...768...33R}).

Assuming coplanar configuration $i_{\rm AB}\simeq 90^\circ$, we can use the above given formulas to
estimate the ratio $A_{\rm LITE}/A_{\rm PD}$ for our best systems listed in Tables~\ref{TabLC} and
\ref{TabLITE}. The result is shown in Fig.~\ref{LITEvsPhysical}. For most cases $P_{\rm AB}$ is
long enough compared to $P_{\rm A}$ or $P_{\rm B}$, and the ratio $A_{\rm LITE}/A_{\rm PD}$ is
safely higher  then unity. However, there are a few exceptions: the system OGLE LMC-ECL-04623 due
to its orbital period $P_{\rm B} \simeq 10.6$~d of the system B, the longest in our sample, while
the outer orbit has a sufficiently short orbital period ($P_{\rm AB}\simeq 4.55$~yr) and high
eccentricity  ($e_{\rm AB}\simeq 0.5$). Some other systems have a ratio of about 8,
while for the rest the LITE contribution significantly dominates the ETV. All systems having
$A_{\rm LITE}/A_{\rm PD}\lesssim 10$ for one or both components are flagged with a star in
Table~\ref{TabLITE}. We note that the Kepler systems analysed by \citet{2013ApJ...768...33R}
often have  PD dominating LITE  mainly because of their small $P_{\rm AB}$ values (on average about five
times smaller than ours). Our dataset and methods do not allow us to detect such small values of outer
periods.

\subsection{OGLE BLG-ECL-145467}

By far the most interesting target from our sample is OGLE BLG-ECL-145467 thanks to  a conjunction
of several reasons: (i) it is rather bright ($12.65$~mag in I filter out of eclipses), (ii) the two
binaries have conveniently short orbital periods of  ($P_{\rm A}\simeq 3.30$~d and $P_{\rm B}\simeq
4.91$~d) and a relatively short period of their mutual motion ($P_{\rm AB}\simeq 1538$~d), and
(iii) there is a well-detached configuration for both binaries with slightly eccentric orbits.
Additionally, the inferred inner orbital periods imply that this system may be representative of an
interesting category of systems captured in the $3:2$ mean motion resonance (see Sect.~\ref{reso}).
The spectral types are probably  B9 for pair A, while  about B9-A0 for pair B. Assuming $i_{\rm
AB}\simeq 90^\circ$ and a total mass of $\simeq 10$~M$_\odot$, we estimate the amplitude of the
physical delay should amount to about $12$\% of light time effect. So, at the zero approximation
the latter dominates, but a fine analysis would profit from a combination of the two effects in an
interpretation of ETVs. All these factors make this object an ideal target of future detailed
study. More photometric data will help to  refine the analysis of ETVs, but it is mainly
spectroscopy that is needed to help determine the mass of all the individual components in this
system. With that information available, we could also further constrain the overall architecture
of the system, such as the mutual inclination of all participating orbits. This is a prerequisite
for a more detailed dynamical study. For instance, the low eccentricity  of  binary B ($\simeq
0.07$) would, in a standard view, be understood as a formation relic, not yet damped  by tidal
circularization. However, as we argue in Sect.~\ref{reso}, it might also have been excited recently
by capture in the $3:2$ resonant state.

\begin{figure}
 \centering
 \includegraphics[width=0.499\textwidth]{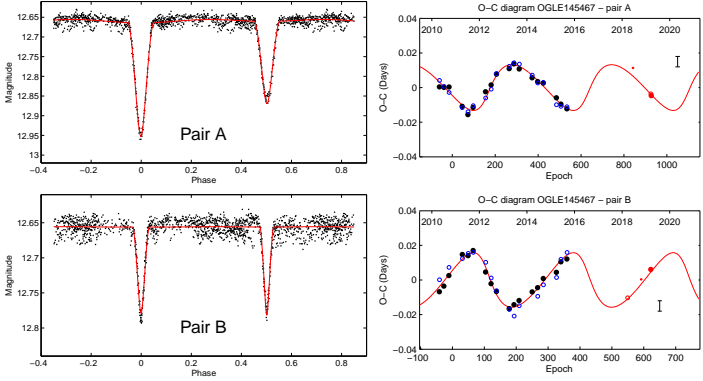}
 \caption{Results of fitting observations of the system OGLE BLG-ECL-145467. Left: Phase-folded light curves of both inner binaries. Right: Period variation (ETVs) as
  the two binaries move on their respective orbits; the primary eclipses are denoted as full dots,
  secondary eclipses as open circles. Shown are our new measurements using the
  Danish 1.54m telescope (red symbols; the size       represents their relative precision). The typical
  uncertainty is plotted as an error bar in the right panels. The red curves are our model.}
 \label{OGLE145467}
\end{figure}

Figure~\ref{OGLE145467} summarizes our modelling  of both  phase-folded LCs of the inner binaries
(left) and  the ETVs revealing their mutual motion about a common centre of mass (right). For
clarity, we subtracted a slow apsidal motion in the right panels;  pair A shows a more prominent
motion with a period of about $125$~yr.  The principal fitted parameters, together with their
uncertainties, are listed in Tables~\ref{TabLC} and \ref{TabLITE}. We note that our own observations include
data over seven nights during the 2018 and 2019 seasons resulting in seven new eclipse epochs,
and suitably complement  data from the OGLE surveys. These data match very closely the ETV trend and
obviously help  to better constrain the  parameters of the mutual orbit. Under the assumption that all
components are located on the main sequence, the results of our modelling imply that the  composite mass of
 binary A is at least $5$~M$_\odot$, while that of   binary B is about $4$~M$_\odot$. Fractional
contributions to the luminosity are $55$\% for pair A and $45$\% for pair B. Given its apparent
magnitude, the system should be at least $2$~kpc  from the Sun. With this solution we can
roughly estimate an angular separation of $\simeq 2.2$~mas of the two binaries for a prospective
interferometric detection.
\begin{table*}
\caption{Parameters derived from the light curve fits of the two inner binaries.}
 \label{TabLC}
 \tiny
  \centering \scalebox{0.69}{
\begin{tabular}{l | c c c c c c c | c c c c c c c}
   \hline\hline\noalign{\smallskip}
 \multicolumn{1}{c|}{System}  &   \multicolumn{7}{c|}{Pair A}                                                                                      &  \multicolumn{7}{c}{Pair B}           \\
 \multicolumn{1}{c|}{name}    & $i_{\rm A}$ [deg] &  $T_1$ [K]    & $T_2$ [K]     & $R_1/a$           & $R_2/a$           & $L_1$ [\%] & $L_2$ [\%] & $i_{\rm B}$ [deg] &  $T_1$ [K]   & $T_2$ [K]     & $R_1/a$           & $R_2/a$           & $L_1$ [\%] & $L_2$ [\%] \\
  \hline\noalign{\smallskip}
 \object{OGLE BLG-ECL-145467} & 86.53 $\pm$ 0.10 & 10700 (fixed) & 8996 $\pm$ 85 & 0.174 $\pm$ 0.002 & 0.154 $\pm$ 0.002 & 64 $\pm$ 1 & 36 $\pm$ 1 & 85.18 $\pm$ 0.11 &10000 (fixed) & 8880 $\pm$ 78 & 0.103 $\pm$ 0.003 & 0.093 $\pm$ 0.004 & 61 $\pm$ 2 & 39 $\pm$ 2 \\ 
 \object{OGLE BLG-ECL-018877} & 86.22 $\pm$ 0.76 &  6000 (fixed) & 6099 $\pm$117 & 0.486 $\pm$ 0.004 & 0.486 $\pm$ 0.004 & 49 $\pm$ 2 & 51 $\pm$ 2 & 76.90 $\pm$ 0.48 & 6000 (fixed) & 5600 $\pm$ 95 & 0.277 $\pm$ 0.008 & 0.224 $\pm$ 0.011 & 65 $\pm$ 5 & 35 $\pm$ 4 \\ 
 \object{OGLE BLG-ECL-061232} & 63.24 $\pm$ 0.74 &  6000 (fixed) & 5816 $\pm$109 & 0.427 $\pm$ 0.005 & 0.427 $\pm$ 0.005 & 52 $\pm$ 2 & 48 $\pm$ 2 & 83.49 $\pm$ 0.43 & 6000 (fixed) & 5629 $\pm$110 & 0.177 $\pm$ 0.010 & 0.175 $\pm$ 0.009 & 53 $\pm$ 3 & 47 $\pm$ 3 \\ 
 \object{OGLE BLG-ECL-088871} & 79.31 $\pm$ 0.15 &  8800 (fixed) & 8525 $\pm$ 73 & 0.202 $\pm$ 0.005 & 0.168 $\pm$ 0.004 & 60 $\pm$ 2 & 40 $\pm$ 1 & 81.22 $\pm$ 0.39 & 7300 (fixed) & 7020 $\pm$ 72 & 0.112 $\pm$ 0.006 & 0.106 $\pm$ 0.009 & 56 $\pm$ 4 & 44 $\pm$ 4 \\ 
 \object{OGLE BLG-ECL-103591} & 84.42 $\pm$ 0.27 &  6350 (fixed) & 6234 $\pm$ 57 & 0.211 $\pm$ 0.003 & 0.152 $\pm$ 0.005 & 67 $\pm$ 3 & 33 $\pm$ 3 & 89.51 $\pm$ 0.95 & 6600 (fixed) & 6701 $\pm$ 84 & 0.190 $\pm$ 0.009 & 0.161 $\pm$ 0.010 & 57 $\pm$ 3 & 43 $\pm$ 3 \\ 
 \object{OGLE LMC-ECL-02903}  & 86.71 $\pm$ 0.29 & 19500 (fixed) &18814 $\pm$156 & 0.219 $\pm$ 0.009 & 0.232 $\pm$ 0.007 & 49 $\pm$ 4 & 51 $\pm$ 4 & 87.70 $\pm$ 0.37 &20500 (fixed) &16210 $\pm$256 & 0.154 $\pm$ 0.004 & 0.078 $\pm$ 0.004 & 85 $\pm$ 2 & 15 $\pm$ 2 \\ 
 \object{OGLE LMC-ECL-04236}  & 78.22 $\pm$ 0.30 & 17000 (fixed) &11540 $\pm$190 & 0.179 $\pm$ 0.005 & 0.141 $\pm$ 0.004 & 74 $\pm$ 3 & 26 $\pm$ 3 & 83.08 $\pm$ 0.92 &11000 (fixed) &10009 $\pm$392 & 0.144 $\pm$ 0.011 & 0.131 $\pm$ 0.010 & 57 $\pm$ 4 & 43 $\pm$ 4 \\ 
 \object{OGLE LMC-ECL-21569}  & 71.32 $\pm$ 0.67 & 35000 (fixed) &23365 $\pm$461 & 0.416 $\pm$ 0.012 & 0.174 $\pm$ 0.010 & 91 $\pm$ 3 &  9 $\pm$ 3 & 87.29 $\pm$ 0.88 &31000 (fixed) &25575 $\pm$608 & 0.223 $\pm$ 0.009 & 0.186 $\pm$ 0.015 & 67 $\pm$ 5 & 33 $\pm$ 4 \\ 
 \object{OGLE BLG-ECL-190427} & 89.50 $\pm$ 0.78 & 31000 (fixed) &15961 $\pm$694 & 0.279 $\pm$ 0.006 & 0.109 $\pm$ 0.004 & 94 $\pm$ 2 &  6 $\pm$ 1 & 85.31 $\pm$ 0.44 &29000 (fixed) &27673 $\pm$503 & 0.164 $\pm$ 0.004 & 0.134 $\pm$ 0.004 & 62 $\pm$ 4 & 38 $\pm$ 3 \\ 
 \object{OGLE GD-ECL-07157}   & 66.74 $\pm$ 0.24 &  7800 (fixed) & 6689 $\pm$ 98 & 0.351 $\pm$ 0.003 & 0.351 $\pm$ 0.003 & 60 $\pm$ 2 & 40 $\pm$ 2 & 89.65 $\pm$ 0.56 & 6500 (fixed) & 6207 $\pm$112 & 0.205 $\pm$ 0.006 & 0.203 $\pm$ 0.009 & 53 $\pm$ 2 & 47 $\pm$ 2 \\ 
 \object{OGLE LMC-ECL-04623}  & 73.55 $\pm$ 0.71 & 22000 (fixed) &17701 $\pm$422 & 0.252 $\pm$ 0.006 & 0.224 $\pm$ 0.005 & 63 $\pm$ 2 & 37 $\pm$ 2 & 87.18 $\pm$ 0.74 &18500 (fixed) &19125 $\pm$940 & 0.125 $\pm$ 0.004 & 0.041 $\pm$ 0.003 & 90 $\pm$ 1 & 10 $\pm$ 2 \\ 
 \object{OGLE LMC-ECL-17182}  & 83.60 $\pm$ 0.59 & 20500 (fixed) & 7663 $\pm$365 & 0.241 $\pm$ 0.011 & 0.097 $\pm$ 0.007 & 96 $\pm$ 1 &  4 $\pm$ 1 & 72.29 $\pm$ 0.53 &20500 (fixed) &10403 $\pm$235 & 0.214 $\pm$ 0.005 & 0.209 $\pm$ 0.004 & 74 $\pm$ 2 & 26 $\pm$ 2 \\ 
 \noalign{\smallskip}\hline                                                                                                                        
\end{tabular}} \\
\end{table*}
\begin{table*}
 \caption{Orbital parameters of the binary orbits, and those of the relative outer
  orbits from ETV analysis assuming the LITE model (see Sect.~\ref{model}).}
 \label{TabLITE}
  \centering \scalebox{0.50}{
\begin{tabular}{l | c c c | c c c | c c c c c c }
   \hline\hline\noalign{\smallskip}
 \multicolumn{1}{c|}{System}     &  \multicolumn{3}{c|}{Pair A}                                           &  \multicolumn{3}{c|}{Pair B}                                           &  \multicolumn{6}{c}{Mutual orbit}   \\
 \multicolumn{1}{c|}{name}       & JD$_0-2450000$ [d]     &  $P_{\rm A}$ [d]          & $e_{\rm A}$       & JD$_0-2450000$ [d]     &  $P_{\rm B}$ [d]          & $e_{\rm B}$       &$P_{\rm AB}$ [yr]& $A_{\rm A}$ [d]   & $A_{\rm B}$ [d]   & $e_{\rm AB}$      & $\omega_{\rm AB}$ [deg]  & $T_0$   \\    
  \hline\noalign{\smallskip}                                                                                                                                                                                                                                                                                %
 \object{OGLE BLG-ECL-145467$^*$}& 5503.1281 $\pm$ 0.0026 & 3.3049105 $\pm$ 0.0000012 & 0.007 $\pm$ 0.001 & 5506.0192 $\pm$ 0.0032 & 4.9097045 $\pm$ 0.0000017 & 0.068 $\pm$ 0.003 & 4.21 $\pm$ 0.24 & 0.013 $\pm$ 0.002 & 0.016 $\pm$ 0.002 & 0.395 $\pm$ 0.018 & 341.2 $\pm$ 9.9 & 2459112 $\pm$ 59 \\    %
 \object{OGLE BLG-ECL-018877}    & 6000.1207 $\pm$ 0.0018 & 0.6008759 $\pm$ 0.0000008 & 0.0               & 6000.3213 $\pm$ 0.0021 & 1.5565025 $\pm$ 0.0000010 & 0.0               & 3.83 $\pm$ 0.14 & 0.012 $\pm$ 0.001 & 0.010 $\pm$ 0.002 & 0.135 $\pm$ 0.032 & 261.6 $\pm$28.0 & 2456089 $\pm$110 \\    %
 \object{OGLE BLG-ECL-061232}    & 6000.2304 $\pm$ 0.0019 & 0.3791298 $\pm$ 0.0000007 & 0.0               & 6000.5497 $\pm$ 0.0024 & 1.4676043 $\pm$ 0.0000018 & 0.0               & 3.42 $\pm$ 0.17 & 0.012 $\pm$ 0.002 & 0.010 $\pm$ 0.003 & 0.451 $\pm$ 0.011 & 265.8 $\pm$14.5 & 2456214 $\pm$142 \\    %
 \object{OGLE BLG-ECL-088871}    & 4504.5283 $\pm$ 0.0041 & 3.8779159 $\pm$ 0.0000024 & 0.0               & 4506.6073 $\pm$ 0.0137 & 5.6508216 $\pm$ 0.0000046 & 0.030 $\pm$ 0.004 &11.27 $\pm$ 3.23 & 0.021 $\pm$ 0.002 & 0.027 $\pm$ 0.003 & 0.349 $\pm$ 0.015 & 248.9 $\pm$ 9.8 & 2456165 $\pm$1123\\    %
 \object{OGLE BLG-ECL-103591}    & 6002.2606 $\pm$ 0.0017 & 2.2321488 $\pm$ 0.0000017 & 0.0               & 6002.9198 $\pm$ 0.0026 & 2.2833714 $\pm$ 0.0000025 & 0.0               & 5.31 $\pm$ 0.44 & 0.011 $\pm$ 0.001 & 0.012 $\pm$ 0.002 & 0.470 $\pm$ 0.031 & 149.0 $\pm$11.3 & 2454137 $\pm$150 \\    %
 \object{OGLE LMC-ECL-02903$^*$} & 5601.1421 $\pm$ 0.0025 & 2.0799677 $\pm$ 0.0000030 & 0.0               & 5604.9207 $\pm$ 0.0110 & 6.5669915 $\pm$ 0.0000301 & 0.260 $\pm$ 0.013 & 4.95 $\pm$ 0.39 & 0.016 $\pm$ 0.001 & 0.015 $\pm$ 0.004 & 0.428 $\pm$ 0.024 & 258.3 $\pm$19.0 & 2456313 $\pm$208 \\    %
 \object{OGLE LMC-ECL-04236}     & 6000.5102 $\pm$ 0.0011 & 2.4074806 $\pm$ 0.0000016 & 0.0               & 6000.6404 $\pm$ 0.0014 & 2.4602955 $\pm$ 0.0000016 & 0.0               & 5.52 $\pm$ 0.11 & 0.005 $\pm$ 0.002 & 0.008 $\pm$ 0.001 & 0.642 $\pm$ 0.050 & 226.7 $\pm$26.3 & 2456711 $\pm$ 49 \\    %
 \object{OGLE LMC-ECL-21569}     & 6004.4492 $\pm$ 0.0032 & 1.9815435 $\pm$ 0.0000047 & 0.0               & 6004.9220 $\pm$ 0.0081 & 2.9328514 $\pm$ 0.0000130 & 0.080 $\pm$ 0.007 &13.82 $\pm$ 1.26 & 0.044 $\pm$ 0.007 & 0.044 $\pm$ 0.010 & 0.316 $\pm$ 0.017 &  15.2 $\pm$ 9.7 & 2458868 $\pm$492 \\    %
 \object{OGLE BLG-ECL-190427}    & 4001.0036 $\pm$ 0.0012 & 0.9449826 $\pm$ 0.0000005 & 0.0               & 4002.3446 $\pm$ 0.0029 & 2.5137669 $\pm$ 0.0000025 & 0.0               & 3.69 $\pm$ 0.30 & 0.003 $\pm$ 0.001 & 0.003 $\pm$ 0.001 & 0.192 $\pm$ 0.047 & 337.6 $\pm$18.6 & 2455650 $\pm$111 \\    %
 \object{OGLE GD-ECL-07157$^*$}  & 2347.4605 $\pm$ 0.0010 & 0.8128751 $\pm$ 0.0000006 & 0.0               & 2348.3299 $\pm$ 0.0210 & 2.6694423 $\pm$ 0.0000590 & 0.101 $\pm$ 0.011 & 5.56 $\pm$ 1.02 & 0.006 $\pm$ 0.002 & 0.008 $\pm$ 0.003 & 0.748 $\pm$ 0.032 & 328.9 $\pm$11.0 & 2457190 $\pm$280 \\    %
 \object{OGLE LMC-ECL-04623$^*$} & 6000.3165 $\pm$ 0.0030 & 1.6422711 $\pm$ 0.0000024 & 0.0               & 6001.8376 $\pm$ 0.0145 &10.6468130 $\pm$ 0.0000816 & 0.0               & 4.55 $\pm$ 0.63 & 0.018 $\pm$ 0.002 & 0.025 $\pm$ 0.009 & 0.501 $\pm$ 0.015 & 207.0 $\pm$10.2 & 2454819 $\pm$177 \\    %
 \object{OGLE LMC-ECL-17182$^*$} & 6000.8695 $\pm$ 0.0068 & 2.2372570 $\pm$ 0.0000042 & 0.120 $\pm$ 0.049 & 6002.3575 $\pm$ 0.0027 & 2.4707908 $\pm$ 0.0000032 & 0.0               & 8.72 $\pm$ 0.17 & 0.015 $\pm$ 0.002 & 0.015 $\pm$ 0.004 & 0.898 $\pm$ 0.049 &  26.2 $\pm$16.1 & 2455632 $\pm$ 88 \\    %
  \noalign{\smallskip}\hline
\end{tabular}} \\
\scriptsize Note: [*] - Systems OGLE BLG-ECL-145467, OGLE LMC-ECL-02903, OGLE GD-ECL-07157, OGLE LMC-ECL-04623, and OGLE LMC-ECL-17182
have a non-negligible contribution of the physical delay
 ($A_{\rm LITE} / A_{\rm PD} < 10$; see Fig.~\ref{LITEvsPhysical}).
\end{table*}
\begin{figure}
 \centering
 \includegraphics[width=0.499\textwidth]{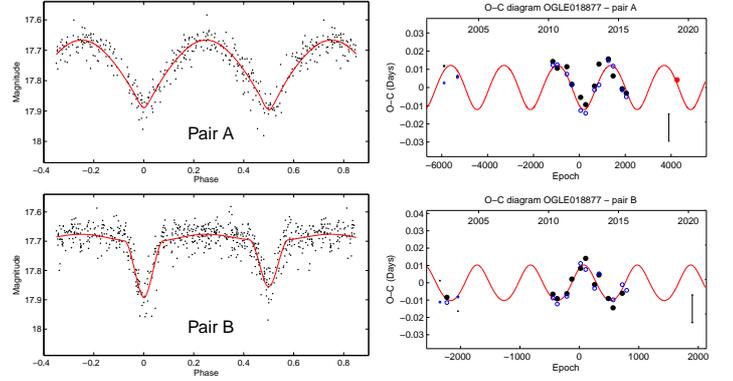}
 \caption{As in Fig.~\ref{OGLE145467}, but for the system
  OGLE BLG-ECL-018877.}
 \label{OGLE018877}
\end{figure}

\subsection{OGLE BLG-ECL-018877}

Another system on our list, OGLE BLG-ECL-018877, is fainter and has a maximum brightness of about
$17.7$~mag in I filter. The inferred surface temperatures (Table~\ref{TabLC})  make us think that all
components are G-type stars. According to our modelling (Fig.~\ref{OGLE018877} and Tables~\ref{TabLC}
and \ref{TabLITE}), the system comprises two compact binaries, one of which is detached and the
other  a  contact binary. Their orbits have undetectable eccentricities with our data. What we call  pair A
(shorter period of $\simeq 0.6$~d) seems to be less massive than  pair B (longer period
of $\simeq 1.56$~d), and their luminosity ratio is about 30\%:70\%. The eccentricity of the mutual
orbit, only about $0.14$, is the lowest in our sample, and  its period, only about $1400$~d,
also belongs to the shortest. Both are somewhat uncertain because of the source faintness, and because the  smaller
amplitude of the ETVs results in a lower signal-to-noise ratio. New accurate observations will
help to constrain these values better.

\subsection{OGLE BLG-ECL-061232}

Similar to OGLE BLG-ECL-018877 is the system OGLE BLG-ECL-061232: it is made up  of a
contact binary A, having a short orbital period of $\simeq 0.38$~d, and a detached binary B with longer
period of $\simeq 1.47$~d, and again the inner eccentricities are not detectable. Additionally, the
mutual motion, unambiguously revealed by the ETVs, has the shortest period in our sample of $\simeq
1250$~d, though the eccentricity is slightly higher. We note that because of very short periods
$P_{\rm A}$ and $P_{\rm B}$, this and the previous systems should have their observed ETVs safely
dominated by the light time effect (see also Fig.~\ref{LITEvsPhysical}). Assuming normal main
sequence components, both binaries consist of F-G spectral type stars, while  binary A slightly
dominates; the luminosity ratio is approximately 65\%:35\%.

\begin{figure}
 \centering
 \includegraphics[width=0.499\textwidth]{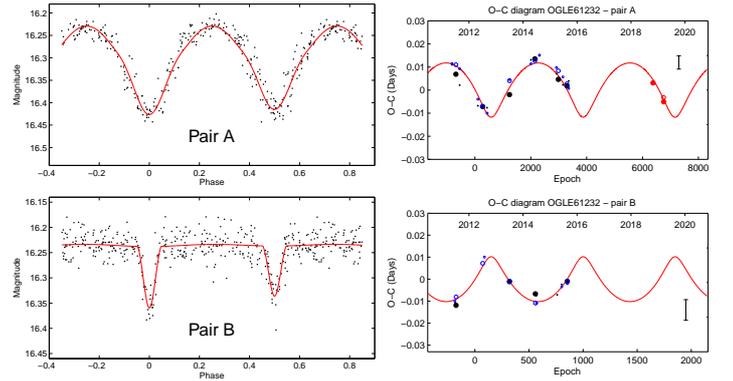}
 \caption{As in Fig.~\ref{OGLE145467}, but for the system
  OGLE BLG-ECL-061232.}
 \label{OGLE061232}
\end{figure}

\subsection{OGLE BLG-ECL-088871}

\begin{figure}
 \centering
 \includegraphics[width=0.499\textwidth]{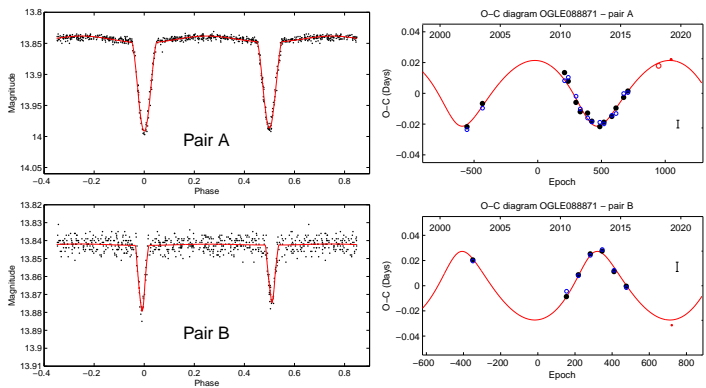}
 \caption{As in Fig.~\ref{OGLE145467}, but for the system
  OGLE BLG-ECL-088871.}
 \label{OGLE088871}
\end{figure}

Another system with a pair of well-detached binaries is OGLE BLG-ECL-088871, an alter-ego of OGLE
BLG-ECL-145467. It is bright enough to have a good prospect in follow-up observations, including
spectroscopy. It consists of wide systems with $P_{\rm A}\simeq 3.88$~d and $P_{\rm B}\simeq
5.65$~d, apparently very close to the $3:2$ resonance. Binary B, with longer period, has a low
eccentricity $e_{\rm B}\simeq 0.03$, but no apsidal motion has been detected in our data. This is
likely because the mutual orbit  has a longer period of $\simeq 4116$~d, implying the period of the
apsidal motion for system B is long and the stars distant enough from each other such that tidal
interaction is weak. The large value of $P_{\rm AB}$ also implies that the physical delay
contributes ETVs almost negligibly (see Fig.~\ref{LITEvsPhysical}). The light curve analysis in
Fig.~\ref{OGLE088871} reveals that the eclipses of binary A are about five times deeper than those
of  binary B, likely an expression of a slight non-coplanarity. Stars in binary B contribute  only
about 35\% of the total luminosity of the system. All the components seem to be A-type stars.

\subsection{OGLE BLG-ECL-103591}

\begin{figure}
 \centering
 \includegraphics[width=0.499\textwidth]{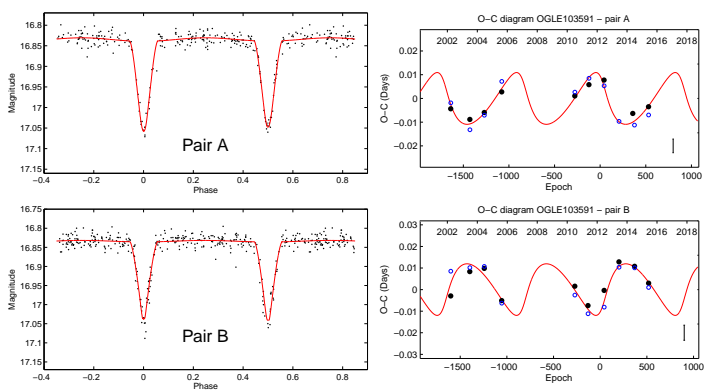}
 \caption{As in Fig.~\ref{OGLE145467}, but for the system
  OGLE BLG-ECL-103591.}
 \label{OGLE103591}
\end{figure}

The analysis of yet another well-detached system OGLE BLG-ECL-103591 is shown in Fig.~\ref{OGLE103591}.
With orbital periods of the inner binaries of $\simeq 2.23$~d and $\simeq 2.28$~d, it is
representative of the $P_{\rm A}\simeq P_{\rm B}$ configurations, though unlikely to be located in
the corresponding resonance (see discussion in Sec.~\ref{reso}). All the components are probably
F type, while the individual relative luminosities of binaries A and B contribute  60\% and
40\% of the total luminosity of the system, in agreement with their ETV amplitudes $A_A$ and
$A_B$.

\subsection{OGLE LMC-ECL-02903}

\begin{figure}
 \centering
 \includegraphics[width=0.499\textwidth]{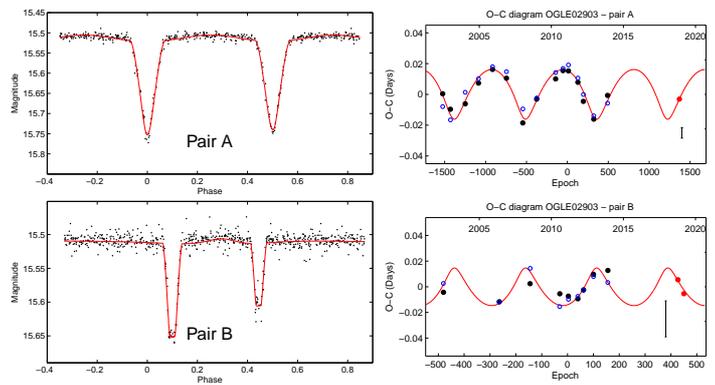}
 \caption{As in Fig.~\ref{OGLE145467}, but for the system
  OGLE LMC-ECL-02903.}
 \label{OGLE_LMC02903}
\end{figure}

Our next example, system OGLE LMC-ECL-02903, is located in the Large Magellanic Cloud. While
consisting of two detached binaries, like several others mentioned earlier, this system is an exception because of the high
eccentricity of   binary B:  $e_{\rm B}\simeq 0.26$ (see Fig.~\ref{OGLE_LMC02903}). Because
the system is not overly compact, the period of mutual revolution of the A and B binaries is
$\simeq 1808$~d, the apsidal motion of   binary B has quite a long period of $\simeq 500$~yr and is
barely detectable in our dataset.   Binary B seems to be slightly more massive than  binary A;
 the corresponding fraction of the whole luminosity is 47\% and 53\%, respectively. The
spectral classification of all components in this quadruple seems to be of B type.

\subsection{OGLE LMC-ECL-04236}

The second system located in the LMC fields is called OGLE LMC-ECL-04236. It is composed of two pairs
of binaries with their respective orbital periods of $2.41$~d and $2.46$~d. However, both binaries
are rather different from each other.  Pair A dominates the system (it produces about 80\% of
the total luminosity) and  probably consists of two B3 stars. The second pair's  components are of
a significantly later spectral type (about B8-B9). Their mutual orbit has a period of about
$2016$~d, but new observations are still needed to better constrain its orbital parameters
because the ETV amplitude is rather small, $\lesssim 0.01$~d, and thanks to the faintness of the system the
signal-to-noise ratio is rather poor.

\begin{figure}
 \centering
 \includegraphics[width=0.499\textwidth]{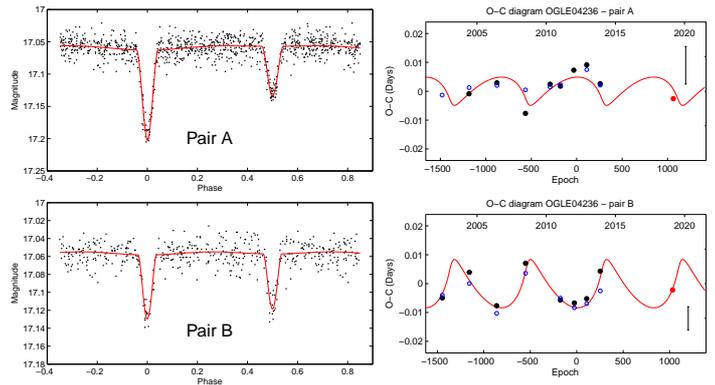}
 \caption{As in Fig.~\ref{OGLE145467}, but for the system
  OGLE LMC-ECL-04236.}
 \label{OGLE_LMC04236}
\end{figure}

\subsection{OGLE LMC-ECL-21569}

\begin{figure}
 \centering
 \includegraphics[width=0.499\textwidth]{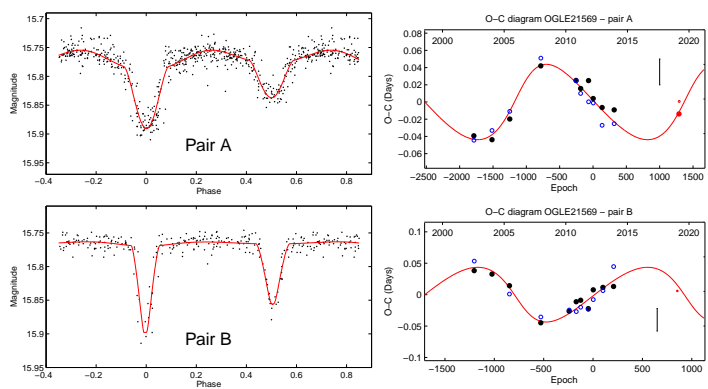}
 \caption{As in Fig.~\ref{OGLE145467}, but for the system
  OGLE LMC-ECL-21569.}
 \label{OGLE_LMC21569}
\end{figure}

OGLE LMC-ECL-21569 is another very interesting system in our portfolio. Photometric analysis reveals
short orbital periods for binaries A and B:  $P_{\rm A}\simeq 1.98$~d and $P_{\rm B}\simeq
2.93$~d. These values are again suspiciously close to the $3:2 $ ratio, perhaps evidence of
residence in the corresponding mean motion resonance. Binary B is notable for its low
eccentricity ($e_{\rm B}\simeq 0.08$), and its rather short  period of  apsidal motion, only
$\simeq 100$~yr. Interestingly, the amplitude of the ETVs is the largest in our sample, and
its mutual orbital period $P_{\rm AB}$ is the longest, nearly $\simeq 5046$~d (luckily, we secured
observations of this interesting target in the 2018-2019 season and this helped  to constrain better
this long $P_{\rm AB}$ period; Fig.~\ref{OGLE_LMC21569}). This implies that  the light time effect
dominates the physical delay (see also Fig.~\ref{LITEvsPhysical}) and the masses of the two binaries
should thus be the same.

OGLE LMC-ECL-21569 is unique among our Group~1 systems because some spectral observations were
published (\citealt{2013A&A...550A.107S}, and \citealt{2017A&A...598A..84A}) and the star is classified as an
O8V object. Unfortunately, only the lines of the most dominant component of pair A were detected with
no signature of the four components, hence we are still dealing with a single-lined SB1 spectroscopic binary. From the amplitudes of the LITEs of both pairs we would expect to also  have  similar luminosity
contributions from both pairs, but this is not true here.
For the analysis we used an assumption of standard mass-luminosity-bolometric magnitude relations
for stars on the main sequence (see tables in e.g.   \citealt{2013ApJS..208....9P} with recent updates
online,\footnote{\tiny{www.pas.rochester.edu/$\sim$emamajek/EEM$\_$dwarf$\_$UBVIJHK$\_$colors$\_$Teff.txt}}
and  a method of estimating the photometric mass ratio by \citealt{2003MNRAS.342.1334G}). From
the radial velocities we obtain $a=20.3~\mathrm{R_\odot}$ and $q_A=0.26$ for pair A, hence
the secondary of A is about B2-3. 
For  pair B the situation is as follows:    $q_B=0.73$ and both B
components are of about B0-1 spectral type. From this it   follows that the luminosity
fraction of the A and B pairs is   about 75\%:25\%. In addition, both components of pair B are less luminous
than the primary of A, which should be the reason why only the lines of the O8V star were detected in
the spectra. This result of our fitting is plotted in Fig. \ref{OGLE_LMC21569} and is also given in
Tables \ref{TabLC} and \ref{TabLITE}.

\subsection{Other systems from Group 1}

\begin{figure*}
 \centering
 \includegraphics[width=0.999\textwidth]{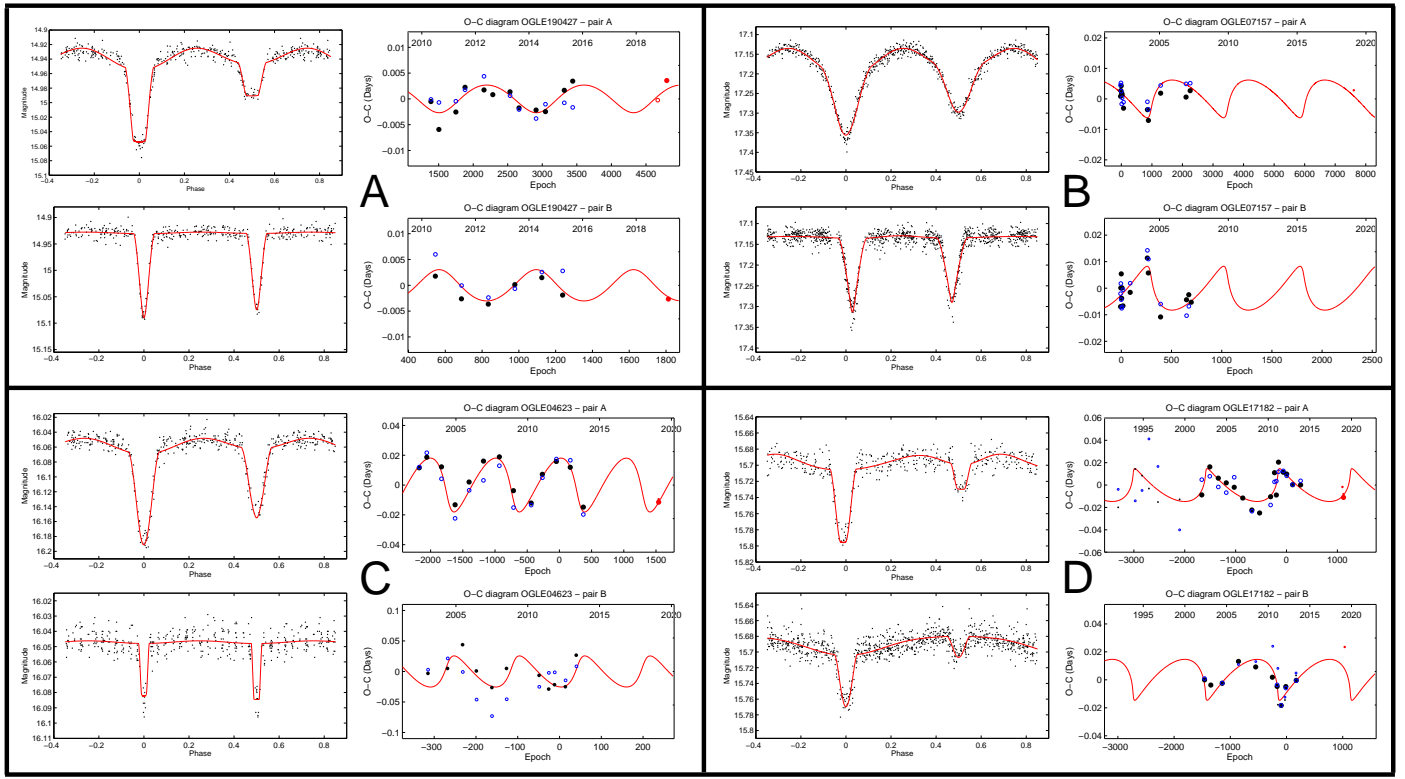}
 \caption{Results of fitting of four other systems from   Group 1 (using the same
  scheme as in Fig.~\ref{OGLE145467}): (A) OGLE BLG-ECL-190427 (top   left), (B)
  OGLE GD-ECL-07157 (top   right), (C) OGLE LMC-ECL-04623 (bottom   left), and
  (D) OGLE LMC-ECL-17182 (bottom   right).}
 \label{OGLE_4systems}
\end{figure*}

 Here we only briefly mention the remaining systems in Group 1, which all have adequately good
coverage of their light curves and also ETVs detectable for both binaries A and B, yet their
analysis should be still considered as preliminary (see Fig.~\ref{OGLE_4systems}). Often the
problem has to do with less than optimum coverage of the relative orbit by ETVs, their low signal-to-noise ratio, and so on. We consider the case of OGLE BLG-ECL-190427, which has the smallest
amplitude of ETVs among our studied quadruples (the amplitude of $\simeq 0.003$~d is just $0.1$\%
of the orbital period of binary B). Binary A in the system OGLE LMC-ECL-17182 exhibits a fast
apsidal motion, an estimated period of only  $\simeq 70$~yr. Similarly,  binary B in the system OGLE
GD-ECL-07157 shows an apsidal motion with a slightly longer period of $\simeq 130$~yr. Both of these
periods are rather uncertain due to fairly incomplete coverage by data. Their uncertain
fit may affect our derived ETV series. Interestingly, both systems also have the highest
eccentricity $e_{\rm AB}$ of the relative orbit.

\subsection{1SWASP J093010.78+533859.5}

Let us finally point out another doubly eclipsing system that does not belong to Group 1 with
well-derived orbit, but  still adequately interesting. We paid attention to  system 1SWASP
J093010.78+533859.5 for several reasons: (i) it is one of only a few systems on the northern sky
(see Sect.~\ref{stat}); (ii) it is quite bright, and can be observed by even
small-scale telescopes; and (iii) it has already been the target of several publications
\citep[see][]{2013A&A...549A..86L,2014AJ....147..104K,2015AaA...578A.103L,2018Ap.....61..458H}. In
spite of this effort, none of the previous studies revealed evidence of the relative motion of the
two binaries having rather short orbital periods $P_{\rm A} \simeq 1.31$~d and $P_{\rm B}\simeq
0.23$~d.

\begin{figure}
 \centering
 \includegraphics[width=0.42\textwidth]{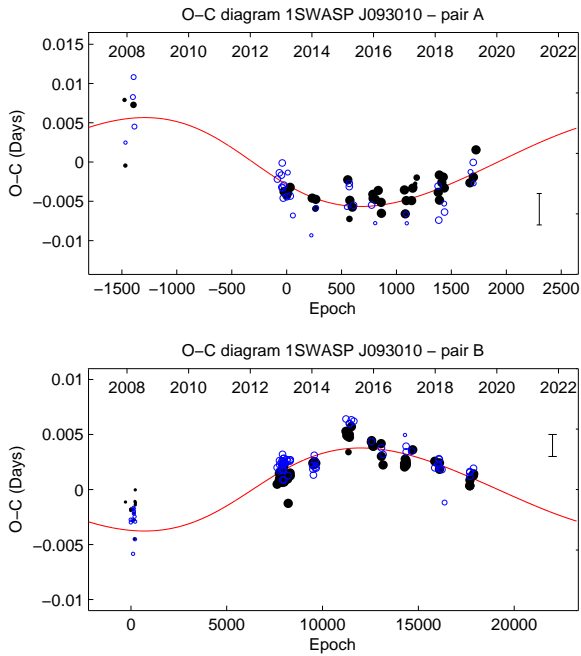}
 \caption{System 1SWASP J093010.78+533859.5 showing the ETV series
  of the eclipsing binaries: A (top) and  B (bottom).}
 \label{TYC3807}
\end{figure}

We applied the same approach as above to the observations of the 1SWASP J093010.78+533859.5 system;
in particular, we constructed and analysed the ETV series for the two binaries. The result, shown
in Fig.~\ref{TYC3807}, indeed suggests a regular signal of the LITE due to relative motion.
However, due to a long period $P_{\rm AB}$, likely longer than $15$~yr, the coverage of the full
cycle of mutual motion is still incomplete. As a result, all the parameters,  such as $e_{\rm
AB}\simeq 0.2$, $A_{\rm A}\simeq 0.006$~d, or $A_{\rm B}\simeq 0.004$~d,  are still quite
uncertain. Long-term follow-up photometric observations during the next decade are thus strongly
encouraged.

Previous studies of this system by \cite{2014AJ....147..104K} and \cite{2015AaA...578A.103L}
allowed us to determine a distance to 1SWASP J093010.78+533859.5 of $66-78$~pc. This is an
interestingly small value. Assuming our solution of $P_{\rm AB}$ is correct, we can estimate
the maximum predicted angular separation of the two binaries on the sky. We obtained a range $\simeq
(90-110)$~mas. This value is comfortably within the detection limit of modern optical
interferometers, and   given  its rather high brightness (about $10$~mag in V filter), it should
be a viable task to resolve the double and derive the relative orbit of the two binaries using
interferometry. This data will provide independent constraints on the physical parameters of all
participating stars in this interesting quadruple, and will allow us to derive the mutual inclination of
the orbits.

\subsection{Other systems from Group 2 and Group 3}

Finally, we make a few comments on several Group 2 and Group 3 systems, namely those for which
the dataset is insufficient for reliable confirmation of LITE variation in their ETVs. Often there
are too few observations to construct meaningful series of ETVs, the noise is too high or
the amplitude is too small, the period $P_{\rm AB}$ of relative orbit is too long, or some  other issue.
Still, some ETVs were detected and many of the systems warrant further observations. Specific
comments on these systems are given in Table~\ref{OtherSystems}.

A peculiar behaviour  has been observed in the system OGLE LMC-ECL-22891, for which the ETVs of
both binary A and B are in phase and behave similarly. This contradicts our simple model based on
LITE and their motion about a common centre of mass. Unless there is a more complicated
explanation, there is always the possibility of a fifth star in the system accompanying as a
distant component of the quadruple consisting of the two eclipsing binaries. Certainly more
observations are needed to clarify the situation.

\begin{table}
 \caption{Some other analysed systems. The upper part comprises  the  systems of Group 2,
  while the lower part of Group 3.} \label{OtherSystems}
\tiny
\begin{tabular}{p{30mm}|p{51mm}}
   \hline\hline\noalign{\smallskip}
  System      & Remark  \\
  \hline\noalign{\smallskip}
 OGLE LMC-ECL-02310  & Too long mutual period $P_{\rm AB}$ ($>$11yr), insufficient data coverage. Long period $P_{\rm B}$. \\
 OGLE LMC-ECL-15607  & Too small LITE variation for binary A, period $P_{\rm AB}$ long ($>$16yr), shallow eclipses of binary B. \\
 OGLE LMC-ECL-15742  & Only very small amplitude of LITE for both binaries (smaller than $0.004$~d), period $P_{\rm AB}$ about $7$~yr. Eccentric orbit of binary B, apsidal motion of about $114$~yr. \\
 OGLE LMC-ECL-20932  & Too long period $P_{\rm AB}$ ($>22$~yr). Shallow eclipses of binary B.\\
 OGLE LMC-ECL-21094  & Long period $P_{\rm AB}$ ($>20$~yr), insufficient data. Eccentric orbit of binary A, apsidal motion $>150$~yr. Suspected inclination change of binary B.\\ 
 OGLE BLG-ECL-030128 & Insufficient data for analysis, mutual period $P_{\rm AB}$ about $20$~yr.\\
 OGLE BLG-ECL-335648 & Too short suspected period $P_{\rm AB}$ ($\simeq 2.4$~yr), not enough data for proper analysis.\\
 OGLE BLG-ECL-398110 & Mutual period $P_{\rm AB}$ of about $7.4$~yr, but insufficient data coverage for detailed analysis. Only shallow eclipses of binary A. \\
 TYC 3929-724-1      & Detailed analysis will be published in a forthcoming separate study. \\
 \hline
 OGLE LMC-ECL-02156  & Inclination change of binary A.\\
 OGLE LMC-ECL-16532  & Too slow LITE variation, not detected for binary B, likely because of its too long orbital period $P_B$ ($\simeq 78$~d). \\
 OGLE LMC-ECL-17913  & ETVs of binary A indicate a possible short period of $\simeq 350$~d, but those of  binary B are too noisy. \\
 OGLE LMC-ECL-21994  & Too long orbital period $P_{\rm AB}$ ($>20$~yr), lack of data.\\
 OGLE LMC-ECL-23000  & New data in contradiction with the older data. Solution possible only with an additional quadratic ephemerides term. \\
 \noalign{\smallskip} \hline
\end{tabular}
\end{table}

%

\begin{figure}
 \centering
 \includegraphics[width=0.49\textwidth]{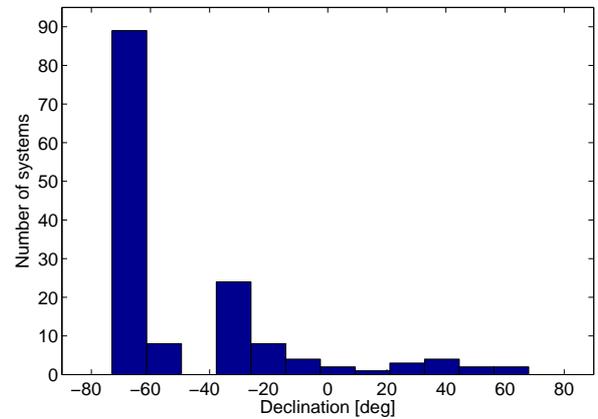}
 \caption{Distribution of currently identified doubly eclipsing systems
  on the sky expressed as a function of their declination. Analysis of data
  provided by OGLE surveys resulted in more than $90$\% of the cases, making them
  located on the southern sky.}
 \label{statistikaDEC}
\end{figure}

\section{Population-scale statistics} \label{stat}

The current sky-distribution of  identified doubly eclipsing systems is very uneven (see
Fig.~\ref{statistikaDEC}), implying that we have information about a small sample of a potentially
much vaster population. The majority of known systems comes from the analysis of data collected by
the OGLE surveys targeting the Magellanic clouds (note the prominent peak at $\simeq -70^\circ$
declination in Fig.~\ref{statistikaDEC}) and the Galactic bulge (note the broader peak between
$\simeq -20^\circ$ and $\simeq -40^\circ$ declination in Fig.~\ref{statistikaDEC}). Long-lasting
and sufficiently homogeneous OGLE data are very fortuitous, thus provided a great opportunity to
search for  doubly eclipsing systems. Only $14$ out of  a total of $146$ of the systems (less than
$10$\%) are located on the northern sky with the largest contribution from the Kepler fields (see
Table~\ref{InfoSystems}). Only a handful of the known systems come from coincidental discoveries.
The sky-distribution skewness has  a second aspect related to the already identified systems. Once
discovered by photometric observations, many of the systems would require a better characterization
by means of spectroscopic or interferometric observations. Oversubscription of the limited number
of telescopes in the southern hemisphere works against this possibility, leaving these interesting
systems poorly understood so far.

\begin{figure}
 \begin{tabular}{c}
  \includegraphics[width=0.47\textwidth]{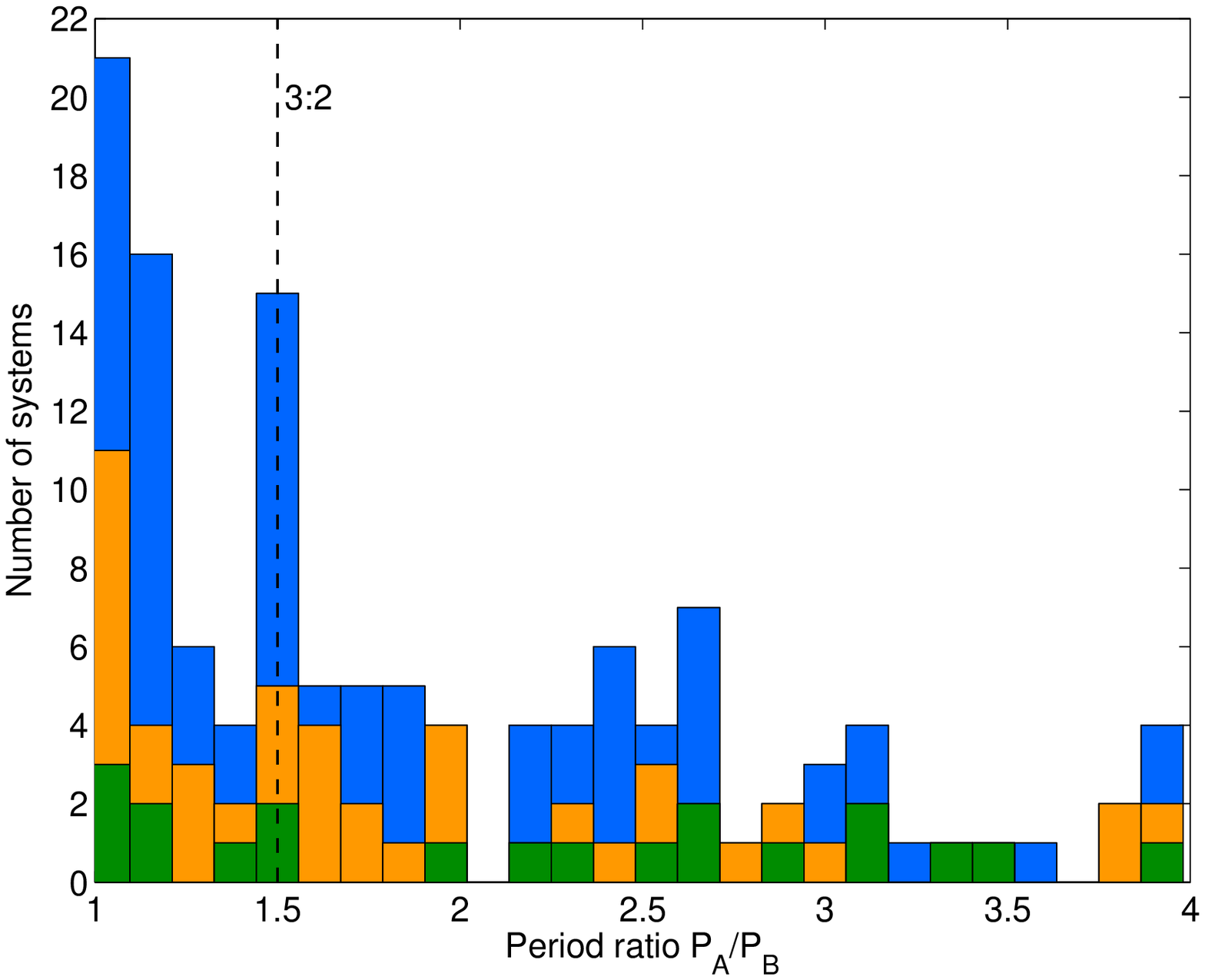} \\
  \includegraphics[width=0.47\textwidth]{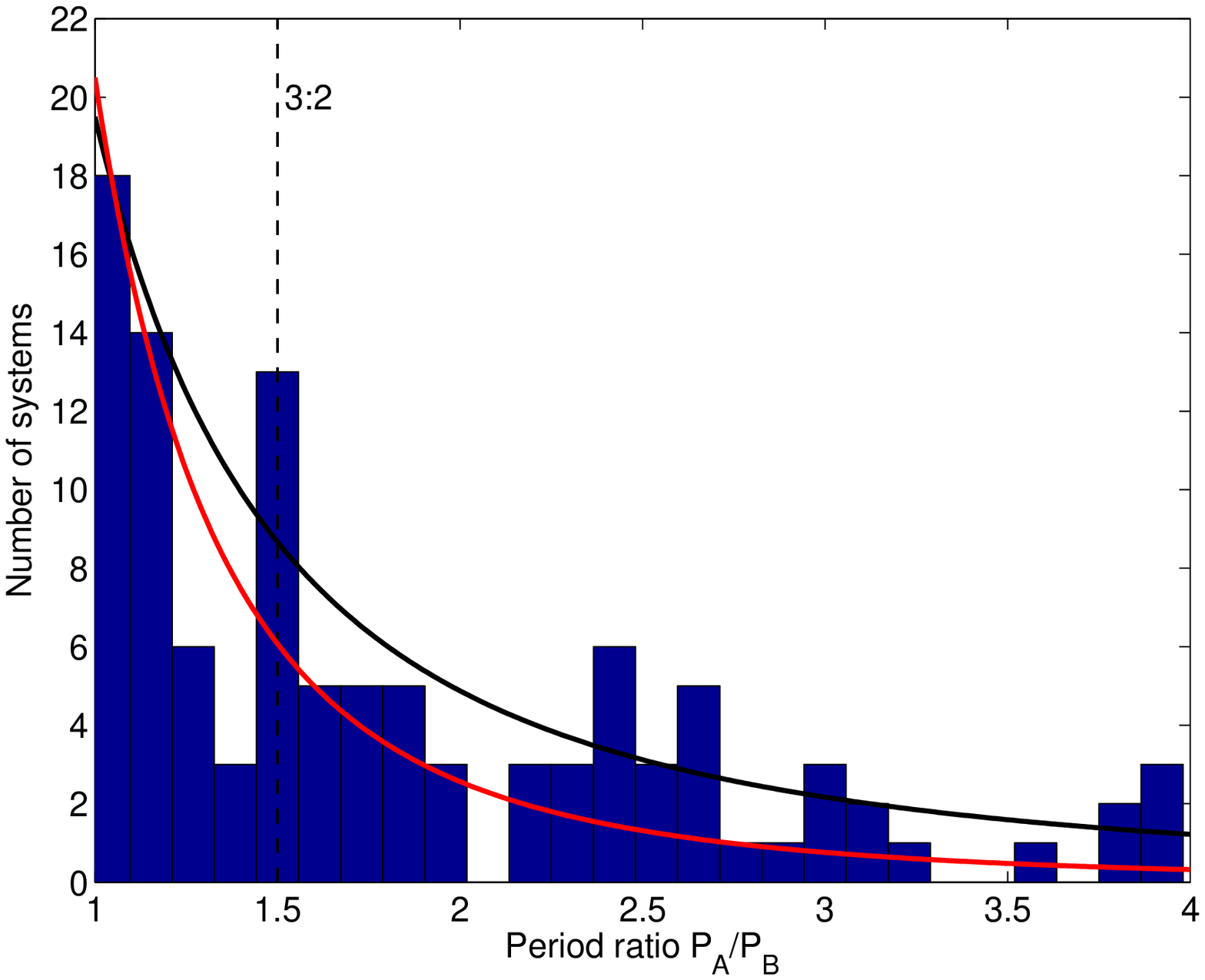} \\
 \end{tabular}
 \caption{Distribution of period ratio $P_{\rm A}/P_{\rm B}$ for binaries in
  quadruple systems (the limiting maximum value of $4$ was chosen because
  information beyond this threshold is very sparse and certainly heavily biased).
  Top panel: All quadruples in the $2+2$ architecture. Blue histograms for previously
  known doubly eclipsing systems, orange histograms for our new discoveries
  (see Table~\ref{InfoSystems}). The data include  the cases where the relative
  motion of the binaries A and B was proven, and those where we lack this information.
  Green histograms are other quadruple systems in the $2+2$ architecture from the MSC catalogue
  \citep{2018ApJS..235....6T}. Bottom panel: Same as above, but only for the
  doubly eclipsing systems (sum of blue and orange above). Declining curves at the
  bottom panel are predictions in a simple model where  periods $P_{\rm A}$ and
  $P_{\rm B}$ are entirely independent variables and have identical probability density
  distribution:  uniform (black) or  linearly increasing towards longer
  orbital periods (red; see discussion in Sect.~\ref{reso}). Both are able to explain the
  preferential $P_{\rm A}\simeq P_{\rm B}$
  configuration. While still noisy, there are two major deviations from their prediction:
  (i) dearth of systems with period ratio between $\simeq 1.2$ and $\simeq 1.5$,
  and (ii) excess of systems with $P_{\rm A}/P_{\rm B}\simeq 3/2$. On the contrary, the
  dip at $P_{\rm A}/P_{\rm B}\simeq 2$ may still reflect data fluctuations.}
 \label{DistributionPerPer}
\end{figure}

Perhaps our most interesting population-scale result is shown in Fig.~\ref{DistributionPerPer},
where we plot the distribution of the orbital period ratio $P_{\rm A}/P_{\rm B}$ for binaries constituting
the 2+2 quadruple systems: (i) for doubly eclipsing systems only (bottom panel), and (ii) for all
systems (upper panel), where the data of doubly eclipsing systems were also incremented by
information of non-eclipsing systems from the MSC catalogue \citep{2018ApJS..235....6T}. In the
case of eclipsing systems, we use also data for those systems where the relative motion has not been proven
yet. By definition, we consider $P_{\rm A}/P_{\rm B}\geq 1$, otherwise we re-index the binaries.
Some systems have $P_{\rm A}/P_{\rm B}>4$, limiting the value shown in Fig.~\ref{DistributionPerPer},
but these data become increasingly sparse. We believe they basically express a growing bias against
the identification of these systems, and thus we disregard this data from our analysis.

Although it is affected by the  still small amount of data, we feel the distribution of $P_{\rm A}/P_{\rm
B}$ reveals several interesting features. Before performing a more involved analysis, and
disregarding bin-to-bin statistical fluctuations, there are  two possible major features: (i)
excess of systems with $P_{\rm A}/P_{\rm B} \simeq 1$, and (ii) excess of systems with $P_{\rm A}/
P_{\rm B} \simeq 1.5$. There is a hint of a similar
feature at $P_{\rm A}/P_{\rm B} \simeq 2.5$, and a dip in the distribution at $P_{\rm A}/P_{\rm B}
\simeq 2$, but they are not statistically robust enough, which leaves us with the two major features.
We believe the interpretation is actually different and we  discuss it briefly in Sect.~\ref{reso}.

Keeping in mind the  underlying dynamical processes, it is interesting to first compare the period
ratio in the $2+2$ quadruples (Fig.~\ref{DistributionPerPer}) with the orbital period ratio of
neighbouring planets. For instance, \citet{2019arXiv190109092Q} in their Fig.~10 show this
information for Kepler candidates in multiplanet systems \citep[see
also][]{2014ApJ...790..146F,pbm19}. There are interesting similarities and dissimilarities.
Starting with the latter, we note that only a few exoplanet configurations are found with a period
ratio below $\simeq 1.2$. This is understood, since the  formation of co-orbiting exoplanets is
possible \citep[e.g.][]{cn06,lyra09,g12}, but it is rather rare. Mutual perturbations tend to
destabilize such configurations \citep[e.g.][]{cn09,r13}. On the contrary, the formation of $2+2$
binaries with near-to-equal orbital periods is easily possible and, as we argue below, perhaps even
preferential. So the major difference in the period-ratio distributions for exoplanets and binaries
in the $2+2$ stellar quadruples readily follows from the orbital architecture.

Focusing now on period ratio values $\gtrsim 1.3$, we note that the distributions for exoplanets and
binaries in quadruples are strikingly similar. Both show a slight decline towards a higher value of
$P_{\rm A}/P_{\rm B}$, punctuated with an excess at resonant value $P_{\rm A}/P_{\rm B} \simeq 1.5$.
The exoplanet data, which are numerous enough, also indicate a dip at $P_{\rm A}/P_{\rm B} \simeq
2$ (followed by an excess at values just wide of this value). There is a hint of such a dip in
the binary data in Fig.~\ref{DistributionPerPer}, but the statistics is still rather poor.  More
data are needed to explore this feature. The excess of planetary orbits at  (or near) the $3:2$
resonant orbital periods is clearly understood by convergent migration followed by resonant capture
\citep[e.g.][]{lw12,bm13b,pbm19}. In the next section, we argue that the same also applies  to the
stellar binary systems in quadruples.


\section{Resonant 2+2 configurations?} \label{reso}

As mentioned above, the statistical distribution of the period ratio $P_{\rm A}/P_{\rm B}$ for our
sample of binaries in the 2+2 quadruples suggests an overabundance of systems with (i) $P_{\rm
A}/P_{\rm B} \simeq 1$ and (ii) $P_{\rm A}/P_{\rm B} \simeq 3/2$, apparently the lowest-order mean
motion resonances $1\!:\!1$ and $3\!:\!2$ between the A and B systems. Interestingly, a careful
analysis of these dynamical configurations has not been carried out yet. This is primarily because
the available observational evidence is poor  and has not motivated their study \citep[as also
mentioned by][these interesting resonances in 2+2 quadruples do not have equivalent configurations
in the planetary systems and therefore have been overlooked in the mainstream orbital
mechanics]{2018MNRAS.475.5215B}. Only recently have \citet{2018MNRAS.475.5215B} given a closer look
to the case of $P_{\rm A}/P_{\rm B} \simeq 1$ resonant configurations, planning to extend their
efforts to the first-order resonances, such as $P_{\rm A}/P_{\rm B}\simeq 2/1$ or $P_{\rm A}/P_{\rm
B}\simeq 3/2$, in their forthcoming work. While still limited, the results from their work helped
us to draw some preliminary conclusions. The point is that the occurrence of systems in these
resonances is only interesting if investigated together with the question of {\it \emph{how they
become resonant in the first place}}. These resonances are very weak and occupy a tiny portion of a
vast space of parameters that characterize  quadruple systems (e.g.  stellar masses, orbital
parameters of their relative orbits). So the most straightforward explanation of their origin is
capture, a process which itself imposes constraints on orbital evolution of the participating
binaries (of tidal or mass-exchange origin).

With regard to the $P_{\rm A}/P_{\rm B} \simeq 1$ configuration, \citet{2018MNRAS.475.5215B}
described in detail the structure of the resonance for planar configuration. They found a
parametric dependence of the resonance width and studied the possibility of  capture.
Interestingly, \citet{2018MNRAS.475.5215B} determined that capture is very unlikely (at least for
the planar systems). Without the option of capture, the number of systems piling up near the first
two bins in the distribution shown in Fig.~\ref{DistributionPerPer} requires a different
explanation. A straightforward possibility is as follows. We assume a simple model in which the
periods of binaries A and B are entirely uncorrelated quantities. Those which contributed to the
construction of data shown in the bottom panel of Fig.~\ref{DistributionPerPer} have a value in a
limited interval, for example $0.5$~d to $15$~d. We assume that both $P_{\rm A}$ and $P_{\rm B}$
have the same probability distribution in this interval. For a uniform distribution, we can easily
show that the ratio $P_{\rm A}/P_{\rm B}$ should have a probability distribution function $\propto
(P_{\rm A}/P_{\rm B})^{-2}$. Likewise, if the probability distribution of $P_{\rm A}$ and $P_{\rm
B}$ linearly increases towards higher values, the ratio $P_{\rm A}/P_{\rm B}$ should have a
probability distribution function $\propto (P_{\rm A}/P_{\rm B})^{-3}$ (and more complex variants
could be easily tested). Each of these predictions (see the declining curves in the bottom panel of
Fig.~\ref{DistributionPerPer}) have a maximum for the equal period situation (i.e. $P_{\rm
A}/P_{\rm B} =1$). This is logical, because there are many more possibilities within the considered
interval of value of $P_{\rm A}$ and $P_{\rm B}$ of this configuration when compared to non-equal
values. Nevertheless, we note that neither of these possibilities matches the observed distribution
of $P_{\rm A}/P_{\rm B}$ exactly. Applying the Kolmogorov-Smirnov test to its cumulative variants,
we obtain that there is only a $\simeq 10^{-5}$ (resp. $\simeq 10^{-4}$) probability of a
statistical match between the data and these simple models (preferring slightly the case with
linearly increasing probability of systems with longer periods, as  might be  visually guessed from
Fig.~\ref{DistributionPerPer}). One of the problems with the predictions is their inability to
describe a dearth of systems with $P_{\rm A}/P_{\rm B}$ in the interval $\simeq 1.2$ to $\simeq
1.5$, prior the significant excess of the $P_{\rm A}/P_{\rm B}\simeq 1.5$ configurations. One
possibility to fix this problem would be to assume a small number of truly $1:1$ dynamically
resonant configurations (perhaps helped by non-coplanar configurations, for which the resonance was
not studied yet). More likely, though, the problem is that the ratio $P_{\rm A}/P_{\rm B}$ is not
static (as implicitly assumed in the previously mentioned, naive model), but time-dependent due to
the tidal interaction between stars in the binaries. We assume, for instance, that the $2+2$
systems preferentially form near the equal-period configuration. If so, their tidal evolution may
be similar and therefore they would remain in the state near the $P_{\rm A}/P_{\rm B} \simeq 1$
ratio for a long time. This would help to increase the excess of this configuration. Modelling the
statistical distribution of $P_{\rm A}/P_{\rm B}$ in such a dynamical model is beyond the scope of
this paper. Nevertheless, its value in describing the population parameters of quadruple systems
should be a motivation to increase number of known systems with the goal  of a less noisy
description of the $P_{\rm A}/P_{\rm B}$ distribution.

None of the above-mentioned simple models also allows us to explain the excess of $P_{\rm A}/P_{\rm
B} \simeq 3/2$ configurations. Here the situation is quite different and we believe the principally
false assumption was the independence of $P_{\rm A}$ and $P_{\rm B}$. This is broken if the
corresponding $3:2$ resonant situation exists. Breiter \& Vokrouhlick\'y (2019, in preparation)
studied the first-order mean-motion resonance in $2+2$ quadruples, again restricting the analysis
to the planar configurations. First, in this case, the resonance strength and width are even
smaller than in the $P_{\rm A}/ P_{\rm B} \simeq 1$ resonance. This is primarily because it
requires non-zero orbital eccentricity $e_{\rm A}$ of the system with longer period $P_{\rm A}$
(for the $3:2$ case), and the resonance strength/width is multiplied by this factor. As an example,
the ratio $P_{\rm A}/P_{\rm B}$ must be within a factor of about $\lesssim 10^{-5}$ near a specific
resonant value. This reference state is not universal, but depends on tidal interaction of stars in
system A (nevertheless, the scatter of the reference values is definitely smaller than the width of
the bins shown in  Fig.~\ref{DistributionPerPer}). Second, the $3:2$ resonance allows capture since
it is adequately described by the second fundamental model of a resonance introduced by
\citet{hl83}. This model has been extensively studied in context of planetary dynamics, including
details of capture conditions \citep[e.g.][]{hl83,eh94}. For instance, if the pre-capture
eccentricity of the A system is very low ($e_{\rm A}\simeq 0$), the capture occurs with a $100$\%
probability and necessarily leads to an increase in $e_{\rm A}$. The capture may be driven by
different effects. Perhaps the most obvious channel relates to decrease in $P_{\rm A}$ and $P_{\rm
B}$  by tidal phenomena. Intriguingly, the theory then requires that $P_{\rm A}$ decreases faster
than $P_{\rm B}$, thus $P_{\rm A}/P_{\rm B}$ approaches $\simeq 1.5$ from an initially higher
value. There are other possible channels of capture, driven for instance by mass exchange or
variations in non-coplanarity of the system, but they have  not been studied yet. Finally, the
capture in the $3:2$ is only temporary and its duration depends on various factors, such as tidal
interaction of stars in the A and B systems. Generally, as the eccentricity $e_{\rm A}$ increases,
the libration amplitude of the critical angle increases as well. Approaching its maximum value of
$180^\circ$, the non-resonant dynamical degrees of freedom lead to a break in the resonant lock and
the system keeps evolving away from the resonance. We note that there are close analogues (but not
exact in details) of behaviour of the first-order resonances in 2+2 quadruples to several problems
in planetary astronomy: (i) capture of dust particles evolving due to the Poynting-Robertston drag
in the exterior resonances with planets \citep[e.g.][]{wj93}, (ii) capture of planetesimals
evolving due to the gas drag in the exterior resonances with planetary embryos
\citep[e.g.][]{wd85}, or (iii) tidally or gas-drag driven dynamics of exoplanets converging towards
their mutual resonances \citep[e.g.][]{lw12,bm13a,bm13b,pbm19}.


Finally, we make a brief comment about the suggested residence of the systems in the resonant state
(especially for the $3:2$ case): Could we prove that those which contribute to the peak in the
$P_{\rm A}/P_{\rm B}$ near $1.5$ value are truly in  resonance? Unfortunately, the answer is no.
From the orbital point of view the resonant state is characterized by the  libration of the critical
angle about the value specified by the stationary point characteristic to the resonance. For
instance, in the case of the $3:2$ resonance the angle $3\lambda_{\rm A}-2\lambda_{\rm
B}-\varpi_{\rm A}$ must librate about the value $180^\circ$ with a small-enough amplitude (here
$\lambda_{\rm A}$ and $\lambda_{\rm B}$ denote longitude in orbit of  systems A and B, and
$\varpi_{\rm A}$ is the longitude of pericentre of  system A). Similarly, in the case of $1:1$
resonance, the critical angle $\lambda_{\rm A}-\lambda_{\rm B}$ must librate either about
$90^\circ$ or $270^\circ$ \citep[e.g.][]{2018MNRAS.475.5215B}. In order to check these properties,
we would need to know accurately enough (i) the whole orbital architecture of the quadruple
system, and (ii) the other physical parameters that define tidal interaction of the stars in
the A and B systems. This is not available at the level needed for any of the systems so far. At
this moment we can only suggest that the systems are either in, or very close to, the resonant state
from the known $P_{\rm A}/P_{\rm B}$ ratio. Interestingly, several of the best characterized
systems suggested to be in the $3:2$ resonance, for which the relative motion of binaries A and B
was proven (e.g. OGLE BLG-ECL-145467, OGLE BLG-ECL-088871, and OGLE LMC-ECL-21569; see
Table~\ref{TabLITE}), exhibit consistent non-zero eccentricity values of the longer-period binary
as required. Further, constraints on the orbital and physical parameters will help in understanding
the dynamical state more accurately.

In conclusion, we note that a detailed census and analysis of resonant 2+2 configurations, in the
first place the $3:2$ case, has a very interesting potential to put constraints on the orbital evolution
of these systems. To achieve this goal, however, the systems must be very carefully characterized.

\section{Conclusions} \label{concl}

We have demonstrated that uniform photometric datasets of long-lasting surveys, such as OGLE in our
case, are very suitable for the analysis of ETV signals in doubly eclipsing quadruple systems. We
succeeded in proving that $28$ of $72$ analysed systems indeed provide evidence of mutual motion of
the two binaries about their common centre of mass, thus constituting gravitationally bound
quadruples. This is  the first time that such a large fraction of new quadruples has been detected.
In most of our cases, we also can interpret the observed ETVs as light-time effect, rather than
physical delay, and this sets a direct estimate of a mass ratio of binaries A and B. Such a
situation is rather unusual in stellar physics, given  that only photometric series are
available.

Furthermore, we may try to draw additional conclusions from the analysis of this limited dataset in
which slightly less than $\simeq 40$\% of systems were positively proven to be gravitationally
bound in this discovery. What about the remaining $\simeq 60$\%? To that end we recall that we
analysed a category of eclipsing binaries with periods of a few days, and the inferred outer
periods were between $\simeq 3$~yr and $\simeq 14$~yr (see Table~\ref{TabLITE}). The architecture of
known triple and quadruple systems, however, easily accommodates outer periods that are  shorter and
especially longer than our restricted interval. Consulting the statistical distribution of long periods
in triples and quadruples by \citet{2008MNRAS.389..925T,2018AJ....155..160T}, for instance, we
estimate that the range of our $P_{\rm AB}$ may represent  $\simeq (10-20)$\% of typical
quadruples. This allows us to believe that  $44$ of the systems we analysed and did not conclusively
determine $P_{\rm AB}$ may well have those values larger, or smaller, than detectable by our
method. This is because our $\simeq 40$\% success rate is actually quite high. It is therefore
conceivable that basically all detected doubly eclipsing systems, namely the $146$ cases listed in
Table~\ref{InfoSystems}, are truly members in gravitationally bound quadruples. With that conclusion
we can also justify that in Sect.~\ref{stat} we might have used all these systems for analysis of
the period ratio $P_{\rm A}/P_{\rm B}$.

The sky location of presently known doubly eclipsing systems in skewed to the southern declinations
for reasons explained in Sec.~\ref{stat} (see Fig. \ref{statistikaDEC}). Interestingly, though, the
VSX database \citep{2006SASS...25...47W} contains information about eclipsing binaries whose number
on northern sky ($\simeq 60000$) basically equals that on the southern sky ($\simeq
80000$). Obviously, the northern systems were compiled from   much less uniform sources if compared
to the southern systems. Nevertheless, we could expect that information about a number of northern
doubly eclipsing systems is already partly available, but they have not been recognized and properly
analysed yet. Still, a dedicated quest for these systems with innovative means of how to use the existing
data could result in further significant increase in the number of known doubly eclipsing
systems.

\begin{acknowledgements}
We would like to thank Andrei Tokovinin for sending us the MSC data for 2+2 systems, and for
his valuable comments about the topic. The research was supported by the grant MSMT INGO II LG
15010. We would also like to thank the Pierre Auger Collaboration for the use of its facilities.
The operation of the robotic telescope FRAM is supported by the grant of the Ministry of Education
of the Czech Republic LM2015038. The data calibration and analysis related to the FRAM telescope is
supported by the Ministry of Education of the Czech Republic MSMT-CR LTT18004 and MSMT/EU funds
CZ.02.1.01/0.0/0.0/16$\_$013/0001402. We also thank the OGLE team for making all of the
observations easily available. We are also grateful to the ESO team at the La Silla Observatory for
their help in maintaining and operating the Danish telescope. This paper makes use of data from the
DR1 of the WASP data \citep{2010A&A...520L..10B}  provided by the WASP consortium, and the
computing and storage facilities at the CERIT Scientific Cloud, reg. no. CZ.1.05/3.2.00/08.0144,
which is operated by Masaryk University, Czech Republic. This paper utilizes public domain data
obtained by the MACHO Project, jointly funded by the US Department of Energy through the University
of California, Lawrence Livermore National Laboratory under contract No. W-7405-Eng-48, by the
National Science Foundation through the Center for Particle Astrophysics of the University of
California under cooperative agreement AST-8809616, and by the Mount Stromlo and Siding Spring
Observatory, part of the Australian National University. The CSS survey is funded by the National
Aeronautics and Space Administration under Grant No. NNG05GF22G issued through the Science Mission
Directorate Near-Earth Objects Observations Program. The CRTS survey is supported by the
U.S.~National Science Foundation under grants AST-0909182 and AST-1313422. This research has made
use of the SIMBAD and VIZIER databases, operated at CDS, Strasbourg, France, and the NASA
Astrophysics Data System Bibliographic Services.
\end{acknowledgements}

\section{Online-only material}

 $\Rightarrow$ Times of eclipses of all systems available upon request!

\end{document}